\documentstyle[12pt]{article}
\input psfig.tex
\newcommand{\be}{\begin{equation}}
\newcommand{\ee}{\end{equation}}

\def \pmbmath{\mathpalette\pmbmathaux}
\def \pmbmathaux#1#2{
         \pmbtext{$#1#2$}}
\def \pmbtext#1{\leavevmode
     \setbox0\hbox{#1}
     \kern-0,2pt \copy0 \kern-\wd0
     \kern0,4pt \copy0 \kern-\wd0
     \kern-0,2pt \raise0,3pt \box0 }

\def\eg{{\it e.g.}\ }

\def\ie{{\it i.e.}\ }

\def\viz{{\it viz.}\ }
\def\vs{{\it vs.}\ }
\def\pa{\parallel}
\def\ep{{\bf \hat{e}}_{\parallel}}
\def\eh{{\bf \hat{e}}}
\def\jp{j^{\prime}}
\def\js{j^{{\prime}{\prime}}}
\def\sp{s^{\prime}}
\def\ss{s^{{\prime}{\prime}}}
\def\kb{{\bf k}}
\def\kbpe{{\bf k}_{\perp}}
\def\kbp{{\bf k}^{\prime}}
\def\kbs{{\bf k}^{{\prime}{\prime}}}
\def\kp{k^{\prime}}
\def\kpe{k_{\perp}}
\def\ks{k^{{\prime}{\prime}}}
\def\kgb{{\pmbmath{\kappa}}}
\def\Lb{{\bf L}}
\def\oh{{{1}\over{2}}}
\def\o4{{{1}\over{4}}}
\def\ph{{{\pi}\over{2}}}
\def\omp{\omega^{\prime}}
\def\oms{\omega^{{\prime}{\prime}}}
\font\tenbi=cmmib10
\newfam\bifam

\textfont\bifam=\tenbi

\begin{document}
\begin{center}
{\bf A WEAK TURBULENCE THEORY FOR INCOMPRESSIBLE MHD} \\
S. Galtier$^{1,2}$, S.\,V. Nazarenko$^1$, A.\,C. Newell$^1$ and A. Pouquet$^2$
\vskip0.4truein
$^1$Mathematics Institute, University of Warwick, Coventry, 
CV4 7AL, U.K.\\
$^2$Observatoire de la C\^ote d'Azur, CNRS UMR 6529, BP 4229, 06304 
Nice--cedex 4, France\\

\vskip0.5truein
{\it J. Plasma Phys., in press}
\end{center}
\vskip0.5truein

%%%%%%%%%%%%%%%%%%%%%%%%%%%%%%%%%%%%%%%%%%%%%%%%%%%%%%%%%%%%%%%%%%%%%%%%%%%%%%%%
\begin{abstract}
We derive a weak turbulence formalism for incompressible MHD.
Three--wave interactions lead to a system of kinetic equations for the spectral
densities of energy and helicity. The kinetic equations conserve energy in all
wavevector planes normal to the applied magnetic field $B_0 \, \ep$. 
Numerically and analytically, we find energy spectra 
$E^\pm \sim k_{\perp}^{n_\pm}$, such that $n_+ + n_- = -4$, where $E^{\pm}$ are 
the spectra of the Els\"asser variables ${\bf z^{\pm}}={\bf v} \pm {\bf b}$ in 
the two-dimensional case ($k_{\pa}=0$). The constants of the spectra are computed 
exactly and found to depend on the amount of correlation between the velocity 
and the magnetic field. Comparison with several numerical simulations and models 
is also made.
\end{abstract}
%%%%%%%%%%%%%%%%%%%%%%%%%%%%%%%%%%%%%%%%%%%%%%%%%%%%%%%%%%%%%%%%%%%%%%%%%%%%%%%%
\vfill\eject

%%%%%%%%%%%%%%%%%%%%%%%%%%%%%%%%%%%%%%%%%%%%%%%%%%%%%%%%%%%%%%%%%%%%%%%%%%%%%%%%
\section{Introduction and General Discussion}
%%%%%%%%%%%%%%%%%%%%%%%%%%%%%%%%%%%%%%%%%%%%%%%%%%%%%%%%%%%%%%%%%%%%%%%%%%%%%%%%

Magnetohydrodynamic (MHD) turbulence plays an important role in many 
astrophysical situations (Parker 1994), ranging from the solar wind
(Marsch and Tu 1994), to the Sun (Priest 1982), the interstellar medium 
(Heiles et al. 1993) and beyond (Zweibel and Heiles 1997), as well as in 
laboratory devices such as tokamaks (see \eg Wild et al. 1981; Taylor 1986;
Gekelman and Pfister 1988; Taylor 1993). A very instrumental step in recognizing 
some of the features that distinguished MHD turbulence from hydrodynamic 
turbulence was taken independently in the early sixties by 
Iroshnikov (1963)
%\cite{iro} 
and Kraichnan (1965)
%\cite{kr65} 
(hereafter IK). They argued that the destruction of phase 
coherence by Alfv\'en waves traveling in opposite directions along local large
eddy magnetic fields introduces a new time scale and a slowing down of energy
transfer to small scales. They pictured the scattering process as being 
principally due to three wave interactions. Assuming 3D isotropy, dimensional 
analysis then leads to the prediction of a $k^{-3/2}$ Kolmogorov finite 
energy flux spectrum. 

However, it is clear, and it has been a concern to Kraichnan and others 
throughout the years, that the assumption of local three dimensional isotropy 
is troublesome. Indeed numerical simulations and experimental measurements both
indicate that the presence of strong magnetic fields make MHD turbulence 
strongly anisotropic. Anisotropy is manifested in a two dimensionalization of
the turbulence spectrum in a plane transverse to the locally dominant magnetic 
field and in inhibiting spectral energy transfer along the direction parallel 
to the field (Montgomery and Turner 1981; Montgomery and Matthaeus 1995; Matthaeus 
et al. 1996; Kinney and McWilliams 1998). Replacing the 3D 
isotropy assumption by a 2D one, and retaining the rest of the IK picture, 
leads to the dimensional analysis prediction of a $k_{\perp}^{-2}$ spectrum 
(${\bf B_0} = B_0 \, \ep$, the applied magnetic field, $k_{\pa} = \kb \cdot \ep$, 
$\kbpe = \kb - k_{\pa} \, \ep$, $k_{\perp}=|\kbpe|$) 
(Goldreich and Sridhar 1997; Ng and Bhattacharjee 1997).

A major controversy in the debate of the universal features of MHD turbulence 
was introduced by Sridhar and Goldreich (1994)
%\cite{srigo} 
(hereafter SG). Following IK,
they assumed that the small-scale MHD turbulence can be described as 
a large ensemble of weakly interacting  Alfv\'en waves within the
framework of the weak turbulence theory. However, SG
challenged that part of IK thinking which viewed Alfv\'en wave scattering as a
three wave interaction process, an assumption implicit in the IK derivation of
the $k^{-3/2}$ spectrum. SG argue that, in the inertial range where amplitudes 
are small, significant energy exchange between Alfv\'en waves can only occur 
for resonant three wave interactions. Moreover, their argument continues, 
because one of the fluctuations in such a resonant triad has zero Alfv\'en 
frequency, the three wave coupling is empty. They conclude therefore that the
long time dynamics of weak MHD fields are determined by four wave resonant 
interactions. 

This conclusion is false. In this paper, we will show that resonant three wave
interactions are non empty (see also Montgomery and Matthaeus 1995; Ng and 
Bhattacharjee 1996) and lead to a relaxation to universal behavior and 
significant spectral energy redistribution. Moreover, weak turbulence theory
provides a set of closed kinetic equations for the long time evolution of the
eight power spectra (to be defined below, in equations (21) and (22)),
corresponding to total energy $e^s(\kb)$, poloidal energy
$\Phi^s(\kb)$, magnetic and pseudo magnetic helicities $R^s(\kb)$, $I^s(\kb)$
constructed from the Els\"asser fields ${\bf z^s} = {\bf v} + s {\bf b}$, 
$s= \pm 1$, where ${\bf v}$ and ${\bf b}$ are the fluctuating velocity and
Alfv\'en velocity respectively. The latter is defined such that ${\bf b} = 
{\bf B}/\sqrt{\mu_0 \rho_0}$, where $\rho_0$ is the uniform density and 
$\mu_0$ the magnetic permeability. 
We will also show that a unique feature of Alfv\'en wave weak 
turbulence is the existence of additional conservation laws. One of the most
important is the conservation of energy on all wavevector planes perpendicular
to the applied field ${\bf B_0}$. There is no energy transfer between planes. 
This extra symmetry means that relaxation to universal behavior only takes place
as function of $k_{\perp}$ so that, in the inertial 
range (or window of transparency), $e^s(\kb) = f(k_{\pa}) k_{\perp}^{-p}$ where 
$f(k_{\pa})$ is non universal. 

Because weak turbulence theory for Alfv\'en waves is not straightforward and 
because of the controversy raised by SG, it is important to discuss carefully
and understand clearly some of the key ideas before outlining the main results. 
We therefore begin by giving an overview of the theory for the statistical 
initial value problem for weakly nonlinear MHD fields.

%%%%%%%%%%%%%%%%%%%%%%%%%%%%%%%%%%%%%%%%%%%%%%%%%%%%%%%%%%%%%%%%%%%%%%%%%%%%%%%%
\subsection{Alfv\'en weak turbulence\,: the kinematics, the asymptotic closure
and some results}
%%%%%%%%%%%%%%%%%%%%%%%%%%%%%%%%%%%%%%%%%%%%%%%%%%%%%%%%%%%%%%%%%%%%%%%%%%%%%%%%

Weak turbulence theory is a widely familiar approach to the plasma physics 
community, see e.g. Vedenov (1967, 1968), Sagdeev and Galeev (1969), Tsytovich 
(1970), Kuznetsov (1972, 1973), Zakharov (1974, 1984), Akhiezer et al. (1975),
McIvor (1977), Achterberg (1979) and Zakharov et al. (1992). 
This approach considers statistical states which can be viewed as large ensembles 
of weakly interacting waves and which can be described by a kinetic equation for 
the wave energy. Recall that the IK theory considers large ensembles of weakly 
interacting Alfv\'en waves, but IK do not derive a kinetic equation and they 
restrict themselves to  phenomenology based on the dimensional argument.
Ng and Bhattacharjee (1996)
%\cite{ngb1} 
developed a theory of weakly interacting Alfv\'en wave packets which 
takes into account anisotropy which leads to certain predictions for the 
turbulence spectra based on some additional phenomenological assumptions and 
by-passing derivation of the weak turbulence kinetic equations. To date, 
there exists no rigorous theory of weak Alfv\'en turbulence in  incompressible MHD,
and derivation of such a theory via a systematic asymptotic expansion in powers of 
small nonlinearity is the main goal of the present paper. It is interesting that 
the kinetic equations were indeed derived (in some limits) for the Alfv\'en waves 
for the cases when such effects, as finite Larmor radius (Mikhailovskii at al. 
1989) or compressibility (Kaburaki and Uchida 1971; Kuznetsov 1973), make these 
waves being dispersive.
Perhaps the main reason why such a theory has not been developed for the 
non-dispersive Alfv\'en waves in incompressible magneto-fluids was a general 
understanding within the "weak turbulence" community that a consistent asymptotic 
expansion is usually impossible for nondispersive waves. The physical reason for 
this is that all wavepackets propagate with the same group velocity even if their 
wavenumbers are different.
Thus, no matter how weak the nonlinearity is, the energy exchanged between the 
wavepackets will be accumulated over a long time and it may not be considered 
small, as it would be required in the weak turbulence theory. As we will show in 
this paper, the Alfv\'en waves represent a unique exception from this rule. This 
arises because the 
nonlinear interaction coefficient for co-propagating waves is null, whereas the 
counter-propagating wavepackets pass through each other in a finite time and 
exchange only small amounts of energy, which makes the weak turbulence approach 
applicable in this case. Because of this property, the theory of weak Alfv\'en 
turbulence which is going to be developed in this paper posesses a novel and 
interesting mathematical structure which is quite different from the classical weak
turbulence theory of dispersive waves.

The starting point in our derivation
is a kinematic description of the fields. We assume that the
Els\"asser fields ${\bf z^s}({\bf x},t)$ are random, homogeneous, zero mean fields
in the three spatial coordinates ${\bf x}$. This means that the n-point 
correlation functions between combinations of these variables estimated at 
${\bf x_1},...,{\bf x_n}$ depend only on the relative geometry of the spatial 
configuration. We also assume that for large separation distances
$|{\bf x_i} - {\bf x_j}|$ along {\bf any} of the three-spatial directions, 
fluctuations are statistically independent. We will also discuss the case of 
strongly two dimensional fields for which there is significant correlation along 
the direction of the applied magnetic field. We choose to use cumulants rather 
than moments, to which the cumulants are related by a one-to-one map. The choice 
is made for two reasons. The first is that they are exactly those combinations of 
moments which are asymptotically zero for all large separations. Therefore they 
have well defined and, at least initially before long distance correlations can 
be built up by nonlinear couplings, smooth Fourier transforms. We will be 
particularly interested in the spectral densities
\be
q^{s s^{\prime}}_{j j^{\prime}}(\kb) = {1 \over (2 \pi)^3} \int_{-\infty}^
{+ \infty} \langle z^s_j({\bf x}) z^{s^{\prime}}_{j^{\prime}}({\bf x} + {\bf r}) 
\rangle \, e^{- i \kb \cdot {\bf r}} \, d{\bf r}
\ee
of the two point correlations. (Remember, $z^{s}_j({\bf x})$ has zero mean so 
that the second order cumulants and moments are the same.) The second reason for 
the choice of cumulants as dependent variables is that, for joint Gaussian 
fields, all cumulants above second order are identically zero. Moreover, because 
of linear wave propagation, initial cumulants of order three and higher decay to 
zero in a time scale $(b_0 k_{\pa})^{-1}$ where ${\bf b_0}={\bf B_0}/\sqrt{\mu_0 
\rho_0}$ is the Alfv\'en velocity ($b_0=|{\bf b_0}|$) and $k_{\pa}^{-1}$ a 
dominant parallel length scale in the initial field. This is a simple 
consequence of the Riemann-Lebesgue lemma\,; all Fourier space cumulants become 
multiplied by fast nonvanishing oscillations because of linear wave properties 
and these oscillations give rise to cancelations upon integration. Therefore, 
the statistics approaches a state of joint Gaussianity. The amount by which it 
differs, and the reason for a nontrivial relaxation of the dynamics, is 
determined by the long time cumulative response generated by nonlinear 
couplings of the waves. The special manner in which third and higher cumulants 
are regenerated by nonlinear processes leads to a natural asymptotic closure of 
the statistical initial value problem.

Basically, because of the quadratic interactions, third order cumulants (equal 
to third order moments) are regenerated by fourth order cumulants and binary 
products of second order ones. But the only long time contributions arise from a 
subset of the second order products which lie on certain resonant manifold 
defined by zero divisors. It is exactly these terms which appear in the kinetic 
equations which describe the evolution of the power spectra of second order 
moments over time scales $(\epsilon^2 \, b_0 k_{\pa})^{-1}$. Here, $\epsilon$ is 
a measure of the strength of the nonlinear coupling. Likewise, higher order 
cumulants are nonlinearly regenerated principally by products of lower order 
cumulants. Some of these small divisor terms contribute to frequency 
renormalization and others contribute to further (\eg four wave resonant 
interactions) corrections of the kinetic equations over longer times. 

What are the resonant manifolds for three wave interactions and, in particular, 
what are they for Alfv\'en waves\,? They are defined by the divisors of a system 
of weakly coupled wave-trains $a_j e^{i(\kb \cdot {\bf x} - \omega_s (k_j)t )}$, 
with $\omega_s (k_j)$ the linear wave frequency, $s$ its level of degeneracy, 
which undergo quadratic coupling. One finds that triads $\kb$, $\kgb$, $\Lb$ 
which lie on the resonant manifold defined for some choice of $s$, $\sp$, $\ss$, 
by 
\be
\begin{array}{lll}
\kb = \kgb + \Lb \, , \\
\omega_s( \kb ) = \omega_{\sp}(\kgb) + \omega_{\ss}(\Lb) \, ,
\end{array}
\label{res}
\ee
interact strongly (cumulatively) over long times $(\epsilon^2 \, \omega_0)^{-1}$, 
$\omega_0$ being a typical frequency. For Alfv\'en waves, $\omega_s ( \kb ) = s 
{\bf b_0} \cdot \kb = s b_0 k_{\pa}$ when $s = \pm 1$ (Alfv\'en
waves of a given wavevector can travel in one of two directions) and $b_0$, the 
Alfv\'en velocity, is the strength of the applied field. Given the dispersion 
relation, $\omega = s {\bf b_0} \cdot \kb$, one might ask why there is any 
weak turbulence for Alfv\'en waves at all because for, $s=\sp=\ss$, (\ref{res}) 
is satisfied for all triads. Furthermore, in that case, the fast oscillations 
multiplying the spectral cumulants of order $N+1$ in the evolution equation for 
the spectral cumulant of order $N$ disappear so that there is no cancelation 
(phase mixing) and therefore no natural asymptotic closure. However, the MHD 
wave equations have the property that the coupling coefficient for this 
interaction is identically zero and therefore the only interactions of 
importance occur between oppositely traveling waves where $\sp=-s$, $\ss=s$. 
In this case, (\ref{res}) becomes 
\be
2s{\bf b_0} \cdot \kgb = 2 s b_0 \kappa_{\pa} = 0 \, .
\ee 
The third wave in the triad interaction is a fluctuation with zero Alfv\'en 
frequency. SG incorrectly conclude that the effective amplitude of this zero mode
is zero and that therefore the resonant three wave interactions are null. 

Although some of the kinetic equations will involve principal value integral 
(PVI) with denominator 
$s \omega(\kb) + s \omega(\kgb) - s \omega(\kb - \kgb) = 2sb_0 \kappa_{\pa}$,
whose meaning we discuss later, the majority of the terms contain the Dirac 
delta functions of this quantity. The equation for the total energy density 
contains only the latter implying that energy exchange takes place by resonant 
interactions. Both the resonant delta functions and PVI arise from taking long 
time limits $t \to \infty$, $\epsilon^2 t$ finite, of integrals of the form 
\be
\begin{array}{lll}
\int F(k_{\pa},\epsilon^2 t) \, \left(e^{(2isb_0k_{\pa}t)} -1 \right) \, 
(2isb_0k_{\pa})^{-1} \, dk_{\pa} \\
\sim \int F(k_{\pa},\epsilon^2 t) \, \left(\pi sgn(t) \, \delta(2sb_0k_{\pa}) + 
i {\cal P} ({1 \over 2sb_0k_{\pa}}) \right) \, dk_{\pa} \, .
\end{array}
\label{int}
\ee
Therefore, implicit in the derivation of the kinetic equations is the assumption 
that $F(k_{\pa},\epsilon^2 t)$ is relatively smooth near $k_{\pa} = 0$ 
so that $F(k_{\pa},\epsilon^2 t)$ remains nearly constant for 
$k_{\pa} \sim \epsilon^2$. In particular, the kinetic equation for the total 
energy density 
\be
e^s({\bf k_{\perp}}, k_{\pa}) = \Sigma_{j=1}^3 q^{s s}_{j j}({\bf k_{\perp}}, 
k_{\pa})
\ee
is the integral over ${{\pmbmath{\kappa}}_{\perp}}$ of a product of a 
combination of 
$q^{s s}_{j j^{\prime}}({\bf k_{\perp}}-{\pmbmath{\kappa}}_{\perp},k_{\pa})$ 
with $Q^{-s}({\pmbmath{\kappa}}_{\perp},0) = \Sigma_{p,m} k_p k_m \, 
q^{-s -s}_{p m} ({\pmbmath{\kappa}}_{\perp},0)$. Three observations (O1,2,3) 
and two questions (Q1,2) arise from this result. 

O1-- Unlike the cases for most systems of dispersive waves, the resonant 
manifolds for Alfv\'en waves {\bf foliate} wavevector space. For typical 
dispersion relations, a wavevector $\kgb$, lying on the resonant manifold 
of the wavevector $\kb$, will itself have a different resonant manifold, and 
members of that resonant manifold will again have different resonant manifolds. 
Indeed the union of all such manifolds will fill $\kb$ space so that energy 
exchange occur throughout all of $\kb$ space.

O2-- In contrast, for Alfv\'en waves, the kinetic equations for the total 
energy density contains $k_{\pa}$ as a parameter which identifies which 
wavevector plane perpendicular to ${\bf B_0}$ we are on. Thus the resonant 
manifolds for all wavevectors of a given ${\bar k_{\pa}}$ is the plane 
$k_{\pa}={\bar k_{\pa}}$. The resonant manifolds foliate $\kb$-space. 

O3-- Further, conservation of total energy holds for each $k_{\pa}$ plane. There
is energy exchange between energy densities having the same $k_{\pa}$ value but
not between those having different $k_{\pa}$ values. Therefore, relaxation 
towards a universal spectrum with constant transverse flux occurs in wavevector 
planes perpendicular to the applied magnetic field. The dependence of the energy
density on $k_{\pa}$ is nonuniversal and is inherited from the initial 
distribution along $k_{\pa}$. 

Q1-- If the kinetic equation describes the evolution of power spectra for values 
of $k_{\pa}$ outside of a band of order $\epsilon^{\xi}$, $\xi < 2$, then how 
does one define the evolution of the quantities contained in 
$Q^{-s}({\pmbmath{\kappa}}_{\perp},0)$ so as to close the system in 
$k_{\pa}$\,?

Q2-- Exactly what is $Q^{-s}({\pmbmath{\kappa}}_{\perp},0)$\,? Could it be 
effectively zero as SG surmise\,?
Could it be possibly singular with singular support located near $k_{\pa}=0$
in which case the limit (\ref{int}) is suspect\,?

To answer the crucially important question 2, we begin by considering the 
simpler example of a one dimensional, stationary random signal $u(t)$ of zero 
mean. Its power spectrum is $f(\omega)$ the limit of the sequence 
$f_L(\omega) = {1 \over 2 \pi} \int^L_{-L} \langle u(t) u(t+\tau) \rangle 
e^{-i\omega \tau} \, d\tau$
 which exists because the integrand decays to zero as $\tau \to \pm \infty$. 
Ergodicity and the stationarity of $u(t)$ allows us to estimate the average 
$R(\tau) = \langle u(t) u(t+\tau) \rangle$ by the biased estimator 
$$R_L(\tau) = {1 \over 2L} \int_{-L+|\tau|/2}^{L-|\tau|/2} u(t-\tau/2) u(t+\tau/2)
\, dt$$
with mean $E\{R_L(\tau)\} = (1-\tau/2L) R(\tau)$. Taking $L$ sufficiently large
and assuming a sufficiently rapid decay so that we can take $R_L(\tau)=0$ for
$|\tau| > 2L$ means that $R_L(\tau)$ is simply the convolution of the signal 
with itself. 
Furthermore the Fourier transform $S_L(\omega)$ can then be evaluated as
$$S_L(\omega) = {1 \over 2\pi} \int_{-2L}^{2L} R_T(\tau) e^{-i \omega \tau} \, 
d\tau = {1 \over 4 \pi L} \, | \int_{-L}^L u(t) e^{-i \omega t} \, dt |^2 \, .$$
For sufficiently large $L$, the expected value of $S_L(\omega)$ is $S(\omega)$, 
the Fourier transform of $R(\tau)$ although the variance of this estimate
is large. Nevertheless $S_L(\omega)$, and in particular $S_L(0)$, is generally 
non zero and measures the power in the low frequency modes. 
To make the connection with Fourier space, we can think of replacing the signal 
$u(t)$ by the periodic extension of the truncated signal 
${\tilde u_L}(t) = u(t)$, $|t| < L$\,; ${\tilde u_L}(t+2L)$ for $|t| > L$. The
zero mode of the Fourier transform $a_L(0) = {1 \over 2L} \int_{-L}^L 
{\tilde u}(t) \, dt$ is a nonzero random variable and, while its expected value
(for large $L$) is zero, the expected value of its square is certainly not zero.
Indeed the expected value of $S_L(0) = 2L \, a_L^2(0)$ has a finite nonzero value
which, as $L \to \infty$, is independent of $L$ as $a_L(0)$ has zero mean and a
standard deviation proportional to $(2L)^{-1/2}$. 
Likewise for Alfv\'en waves, the power associated 
with the zero mode $Q^{-s}({\pmbmath{\kappa}}_{\perp},0)$ is nonzero and 
furthermore, for the class of three dimensional fields in which correlations 
decay in all directions, $Q^{-s}({\pmbmath{\kappa}}_{\perp},k_{\pa})$ is smooth
near $k_{\pa}=0$. Therefore, for these fields, we may consider 
$Q^{-s}({\pmbmath{\kappa}}_{\perp},0)$ as a limit of 
$Q^{-s}({\pmbmath{\kappa}}_{\perp},k_{\pa})$ as 
${k_{\pa} \over k_{\perp 0}} \to 0$ and 
${1 \over \epsilon^2}{k_{\pa} \over k_{\perp 0}} \to \infty$.
Here $k_{\perp 0}$ is some wavenumber near the energy containing part of the 
inertial range. Therefore, in this case, we solve first the nonlinear kinetic 
equation for $\lim_{k_{\pa} \to 0}$ 
$e^s_k(k_{\perp},k_{\pa})$, namely for very oblique Alfv\'en waves, and 
having found the asymptotic time behavior of $e^s(k_{\perp},0)$, then 
return to solve the equation for $e^s(k_{\perp},k_{\pa})$ for finite $k_{\pa}$.

Assuming isotropy in the transverse $k_{\perp}$ plane, we find universal spectra 
$c_n^s k_{\perp}^{n_s}$ for $E^s(k_{\perp})$ 
($\int E^s(k_{\perp},0) \, dk_{\perp} = \int e^s(k_{\perp}) \, d{\bf k_{\perp}}$),
corresponding to finite fluxes of energy from low to high transverse 
wavenumbers. Then $e^s(k_{\perp},k_{\pa}) = f^s(k_{\pa}) c_n^s 
k_{\perp}^{n_s-1}$ where $f^s(k_{\pa})$ is not universal. These solutions each 
correspond to energy conservation. We find that convergence of all integrals is
guaranteed for $-3<n_{\pm}<-1$ and that
\be
n_+ + n_- = - 4
\label{nn}
\ee
which means, that for no directional preference, $n_+ = n_- = -2$. These solutions
have finite energy, \ie $\int E \, dk_{\perp}$ converges. If we interpret them as 
being set up by a constant flux of energy from a source at low $k_{\perp}$ to a 
sink at high $k_{\perp}$, then, since they have finite capacity and can only 
absorb
a finite amount of energy, they must be set up in finite time. When we searched
numerically for the evolution of initial states to the final state, we found a 
remarkable result which we yet do not fully understand. Each $E^s(k_{\perp})$ 
behaves as a propagating front in the form 
$E^s(k_{\perp}) = (t_0-t)^{1/2} E_0(k_{\perp} (t_0-t)^{3/2})$ and 
$E_0(l) \sim l^{-7/3}$ as $l \to +\infty$. This means that for $t<t_0$, the 
$E^s(k_{\perp})$ spectrum had a tail for $k_{\perp} < (t_0-t)^{-3/2}$ with 
stationary form $k_{\perp}^{-7/3}$ joined to $k_{\perp}=0$ through a front 
$E_0(k_{\perp}(t_0-t)^{3/2})$. The $7/3$ spectrum is steeper than the $+2$ 
spectrum. Amazingly, as $t$ approached very closely to $t_0$, disturbances in the
high $k_{\perp}$ part of the $k_{\perp}^{-7/3}$ solution propagated back along 
the spectrum, 
rapidly turning it into the finite energy flux spectrum $k_{\perp}^{-2}$. We 
neither understand the origin nor the nature of this transition solution, nor 
do we understand the conservation law involved with the second equilibrium 
solution of the kinetic equations. Once the connection to infinity is made, 
however the circuit between source and sink is closed and the finite flux energy 
spectrum takes over. 

To this point we have explained how MHD turbulent fields for which correlations
decay in all directions relax to quasiuniversal spectra via the scattering of
high frequency Alfv\'en waves with very oblique, low frequency ones. But there is
another class of fields that it is also important to consider. There are 
homogeneous, zero mean random fields which have the anisotropic property that 
correlations in the direction of applied magnetic field do not decay with 
increasing separation ${\bf B_0} \cdot ({\bf x_1} - {\bf x_2})$. For this case, 
we may think of decomposing the Els\"asser fields as 
\be
z_j^s({\bf x_{\perp}},x_{\pa}) = {\bar z_j^s}({\bf x_{\perp}}) + 
{\hat z_j^s}({\bf x_{\perp}},x_{\pa})
\ee
where the ${\hat z}_j^s({\bf x_{\perp}},x_{\pa})$ have the same properties of the 
fields considered heretofore but where the average of 
$z_j^s({\bf x_{\perp}},x_{\pa})$ 
over $x_{\pa}$ is nonzero. The total average of $z^s_j$ is still zero when one 
averages ${\bar z_j^s}({\bf x_{\perp}})$ over ${\bf x_{\perp}}$. In this case, it 
is not hard to show that correlations 
$$\langle z^s_j ({\bf x_{\perp}},x_{\pa}) \, z^{s^{\prime}}_{j^{\prime}}
({\bf x_{\perp}}+{\bf r_{\perp}},x_{\pa}+r_{\pa}) \rangle$$ 
divide into two parts
$$\langle {\bar z^s_j} ({\bf x_{\perp}}) \, {\bar z^{s^{\prime}}_{j^{\prime}}}
({\bf x_{\perp}}+{\bf r_{\perp}}) \rangle + 
\langle {\hat z}^s_j({\bf x_{\perp}},x_{\pa}) \, 
{\hat z}^{s^{\prime}}_{j^{\prime}}
({\bf x_{\perp}}+{\bf r_{\perp}},x_{\pa}+r_{\pa}) \rangle$$ 
with Fourier transforms,
\be
q_{j j^{\prime}}^{s s^{\prime}}(\kb) = \delta(k_{\pa}) \, 
{\bar q}_{j j^{\prime}}^{s s^{\prime}}(k_{\perp}) + 
{\hat q}_{j j^{\prime}}^{s s^{\prime}}(k_{\perp},k_{\pa}) \, ,
\ee
when ${\bar q}$ is smooth in ${\bf k_{\perp}}$ and ${\hat q}$ is smooth in both 
${\bf k_{\perp}}$ and $k_{\pa}$. The former is simply the transverse Fourier of 
the two point correlations of the $x_{\pa}$ averaged field. Likewise all higher 
order cumulants have delta function multipliers $\delta(k_{\pa})$ for each $\kb$ 
dependence. For example 
$${\bar q}_{j j^{\prime} j^{\prime \prime}}^{s s^{\prime} s^{\prime \prime}}
(\kb,\kb^{\prime}) = \delta(k_{\pa}) \, \delta(k_{\pa}^{\prime}) \, 
{\bar q}_{j j^{\prime} j^{\prime \prime}}^{s s^{\prime} s^{\prime \prime}}
({\bf k_{\perp}},{\bf k_{\perp}^{\prime}})$$
is the Fourier transform of
$$\langle {\bar z}^s_j({\bf x_{\perp}}) \, 
{\bar z}^{s^{\prime}}_{j^{\prime}}
({\bf x_{\perp}}+{\bf r_{\perp}}) \, 
{\bar z}^{s^{\prime \prime}}_{j^{\prime \prime}}
({\bf x_{\perp}}+{\bf r_{\perp}^{\prime}}) \rangle \, .$$

Such singular dependence of the Fourier space cumulants has a dramatic 
effect on the dynamics especially since the singularity is supported precisely on
the resonant manifold. Indeed the hierarchy of cumulant equations for 
${\bar q}^{(n)}$ simply loses the fast (Alfv\'en) time dependence altogether and 
becomes fully nonlinear MHD turbulence in two dimensions with time $t$ replaced 
by $\epsilon t$. Let us imagine, then, that the initial fields are dominated by 
this two dimensional component and that the fields have relaxed on the time 
scale $t \sim \epsilon^{-1}$ to their equilibrium solutions of finite energy 
flux for which ${\bar E}(k_{\perp})$ is the initial Kolmogorov finite energy flux
spectrum $k_{\perp}^{-5/3}$ for $k_{\perp} > k_0$, $k_0$ some input wavenumber 
and ${\bar E}(k_{\perp}) \sim k_{\perp}^{-1/3}$ corresponding to the inverse 
flux of the squared magnetic vector potential (${\bf {\cal A}}\,; {\bf b} = 
\nabla \times {\bf {\cal A}}$). ${\hat {\cal A}}(\kb)$, the  spectral density of 
$\langle {\cal A}^2 \rangle$, behaves as $k_{\perp}^{-7/3}$. These are predicted 
from phenomenological arguments and supported by numerical simulations. 

Let us then ask\,: how do Alfv\'en waves (Bragg) scatter off this two dimensional 
turbulent field\,? To answer this question, one should of course redo all the 
analysis taking proper account of the $\delta(k_{\pa})$ factors in 
${\bar q}^{(n)}$. However, there is a simpler way. Let us imagine that the power 
spectra for the ${\hat z}_j^s$ fields are supported at finite $k_{\pa}$ and 
have much smaller integrated power over an interval $0 \le k_{\pa} < \beta \ll 1$ 
than do the two dimensional fields. Let us replace the $\delta(k_{\pa})$ 
multiplying $q_{j j^{\prime}}^{s s^{\prime}}({\bf k_{\perp}})$ by a function of 
finite width $\beta$ and height $\beta^{-1}$. Then the kinetic equation is 
linear and describes how the power spectra, and in particular ${\hat e}^s(\kb)$, 
of the ${\hat z}^s_j$ fields interact with the power spectra of the two 
dimensional field ${\bar z}^s_j$. Namely, the $Q^{-s}({\bf k_{\perp}},0)$
field in the kinetic equation is determined by the two dimensional field and 
taken as known. The time scale of the interaction is now $\beta \epsilon^{-2}$,
because the strength of the interaction is increased by $\beta^{-1}$ and, is 
faster than that of pure Alfv\'en wave scattering. But the equilibrium of the 
kinetic equation will retain the property that $n_{-s} + n_s = -4$ where now 
$n_{-s}$ is the phenomenological exponent associated with two dimensional MHD
turbulence and $n_s$ the exponent of the Alfv\'en waves. Note that when 
$n_{-s}=-5/3$, $n_s$ is $-7/3$, which is the same exponent (perhaps accidentally)
as for the temporary spectrum observed in the finite time transition to the 
$k_{\perp}^{-2}$ spectrum. 

We now proceed to a detailed presentation of our results.

%%%%%%%%%%%%%%%%%%%%%%%%%%%%%%%%%%%%%%%%%%%%%%%%%%%%%%%%%%%%%%%%%%%%%%%%%%%%%%%%
\section{The derivation of the kinetic equations}
\label{swt2}
%%%%%%%%%%%%%%%%%%%%%%%%%%%%%%%%%%%%%%%%%%%%%%%%%%%%%%%%%%%%%%%%%%%%%%%%%%%%%%%%

The purpose in this section is to obtain closed equations for the energy and 
helicity spectra of weak MHD turbulence, using the fact that, in the presence of 
a strong uniform magnetic field, only Alfv\'en waves of opposite polarities 
propagating in opposite directions interact.

%%%%%%%%%%%%%%%%%%%%%%%%%%%%%%%%%%%%%%%%%%%%%%%%%%%%%%%%%%%%%%%%%%%%%%%%%%%%%%%%
\subsection{The basic equations}
%%%%%%%%%%%%%%%%%%%%%%%%%%%%%%%%%%%%%%%%%%%%%%%%%%%%%%%%%%%%%%%%%%%%%%%%%%%%%%%%

We will use the weak turbulence approach, the ideas of which are described in
great detail in the book of 
Zakharov et al. (1992).
%\cite{ZLF}.
There
are several different ways to derive the weak turbulence kinetic equations. 
We follow here the technique that can be found in 
Benney and Newell (1969). 
%\cite{BN}. 
We write the 3D incompressible MHD equations for the velocity ${\bf v}$ and 
the Alfv\'en velocity ${\bf b}$ 
\be
(\partial_t+{\bf v} \cdot \nabla) \, {\bf v} = - \nabla P_* + 
{\bf b} \cdot \nabla \, {\bf b} + \nu \nabla^2 {\bf v} \, ,
\ee
\be
(\partial_t+{\bf v} \cdot \nabla) \, {\bf b} =
{\bf b} \cdot \nabla {\bf v} + \eta \nabla^2 {\bf b} \, ,
\label{mhd}
\ee
where $P_*$ is the total pressure, $\nu$ the viscosity, $\eta$ the magnetic 
diffusivity and $\nabla \cdot {\bf v}=0$, $\nabla \cdot {\bf b}=0$.
%(with $\mu_0=4\pi \times 10^{-7}$ in free space).
In the absence of dissipation, these equations have three quadratic invariants
in dimension three, namely the total energy 
$E^T={{1}\over{2}} \langle v^2+b^2 \rangle$, the
cross--correlation $E^C=\langle {\bf v}\cdot {\bf b} \rangle$ and the magnetic 
helicity $H^M=\langle {\bf {\cal A}}\cdot {\bf b} \rangle$ (Woltjer 1958).

The Els\"asser variables ${\bf z}^s={\bf v} +s {\bf b}$ with $s=\pm1$
give these equations a more symmetrized form, namely\,: 
\be
(\partial_t+{\bf z}^{-s}\cdot \nabla) \, {\bf z}^s = - \nabla P_* \ ,
\label{mz}
\ee
where we have dropped the dissipative terms which pose no particular closure 
problems. The first two invariants are then simply written as
$2E^s=\langle |{\bf z}^s|^2\rangle$.

We now assume that there is a strong uniform magnetic induction field ${\bf B_0}$
along the unit vector ${\bf \hat{e}}_{\pa}$ and non dimensionalize the equations 
with the corresponding magnetic induction ${\bf B_0}$, where the $z^s$ fields 
have an amplitude proportional to $\epsilon$ ($\epsilon \ll 1$) assumed small 
compared to $b_0$. Linearizing the equations leads to 
\be
(\partial_t - s b_0 \partial_{\pa} ) z^s_j = - \epsilon \partial_{{x_m}} 
z^{-s}_m z^s_j - \partial_{x_j} P_* \, ,
\label{eql}\ee
where $\partial_{\parallel}$ is the derivative along ${\bf \hat{e}}_{\pa}$. 
The frequency of the modes at a wavevector $\kb$ is $\omega (\kb)=\omega_k = 
{\bf b}_0\cdot \kb = b_0 k_{\pa}$. 
We Fourier transform the wave fields using the interaction representation,
\be
z^s_j({\bf x},t) = \int A^s_j(\kb,t) \, e^{i\kb\cdot{\bf x}} \, d\kb 
= \int a^s_j(\kb,t) \, e^{i(\kb\cdot{\bf x}+s \omega_k t)} \, d\kb \, ,
\label{FI}\ee
where $a^s_j(\kb,t)$ varies slowly in time because of the weak nonlinearities\,;
hence 
\be
\partial_t a^s_j(\kb,t)=-i \epsilon k_mP_{jn}\int a^{-s}_m(\kgb) \, 
a^s_n({\bf L}) \, e^{i(-s\omega_k -s\omega_{\kappa} + s\omega_L)t} 
\delta_{\kb, \kgb \Lb} \, d_{\kgb \Lb}
\label{eqa}\ee
with $d_{{\kgb}\Lb}=d{\kgb}d\Lb$ and 
$\delta_{\kb, \kgb \Lb}=\delta(\kb -\kgb - \Lb)$\,; finally,
$P_{jn}(k)=\delta_{jn}-k_jk_nk^{-2}$ is the usual projection operator keeping
the ${\bf A^s}({\bf k})$ fields transverse to ${\bf k}$ because of 
incompressibility.
The exponentially oscillating term in (\ref{eqa}) is essential\,: its exponent
should not vanish when $({\bf k} - {\kgb} - {\bf L})={\bf 0}$, \ie the waves
should be dispersive for the closure procedure to work. In that sense, 
incompressible MHD can be coined ``pseudo''--dispersive
because although $\omega_k\sim k$, the fact that waves of one s--polarity 
interact {\sl only} with the opposite polarity has the consequence 
that the oscillating factor is non--zero except at resonance\,; indeed
with $\omega_k = b_0 k_{\pa}$, one immediately sees that 
$-s\omega_k -s\omega_{\kappa} +s\omega_L=
s(-k_{\pa}-\kappa_{\pa}+L_{\pa})=-2s\kappa_{\pa} $ using the 
convolution constraint between the three waves in interaction. 
In fact, Alfv\'en waves may have a particularly weak form of interactions 
since such interactions take place only when
two waves propagating in opposite directions along the lines of the uniform
magnetic field meet. As will be seen later (see \S \ref{pg}), this has the 
consequence that the transfer in the direction parallel to ${\bf B}_0$
is zero, rendering the dynamics two--dimensional, as is well known
(see \eg Montgomery and Turner 1981; Shebalin et al. 1983). 
Technically, we note that there are two types of waves that 
propagate in opposite directions, so that the classical criterion 
(Zakharov et al. 1992) for resonance to occur, \viz $\omega''>0$ does not apply 
here.

%%%%%%%%%%%%%%%%%%%%%%%%%%%%%%%%%%%%%%%%%%%%%%%%%%%%%%%%%%%%%%%%%%%%%%%%%%%%%%%%
\subsection{Toroidal and poloidal fields}
%%%%%%%%%%%%%%%%%%%%%%%%%%%%%%%%%%%%%%%%%%%%%%%%%%%%%%%%%%%%%%%%%%%%%%%%%%%%%%%%

The divergence--free condition implies that only two scalar fields
are needed to describe the dynamics\,; following classical works in 
anisotropic turbulence, they are taken as (Craya 1958; Herring 1974; Riley et 
al. 1981)
\be
{\bf z}^s={\bf z}^s_1+{\bf z}^s_2 = \nabla\times (\psi^s \ep) + 
\nabla \times (\nabla\times (\phi^s \ep)) \, ,
\label{pp}
\ee
which in Fourier space gives 
\be
A_j^s(\kb) = i \kb \times \ep \, \hat{\psi^s}(\kb) - 
\kb \times ({\bf k} \times \ep) \, \hat{\phi^s}(\kb) \ .
\label{ppf}
\ee
We elaborate somewhat on the significance of the $\hat{\psi^s}$ and 
$\hat{\phi^s}$ fields since they are the basic fields with which we shall deal.
Note that ${\bf z}^s_1$ are two--dimensional fields with no parallel component 
and thus with only a vertical vorticity component (vertical means parallel to
${\bf B_0}$), whereas the ${\bf z}^s_2$ fields have zero vertical 
vorticity\,; such a decomposition is used as well for stratified flows 
(see Lesieur 1990 and references therein). Indeed, rewriting the 
double cross product in (\ref{ppf}) leads to\,:
\be
{\bf A}^s(\kb)=i \kb\times \ep \, \hat{\psi^s}(\kb) - \kb \, k_{\parallel} 
\hat{\phi^s}(\kb) + \ep \, k^2 \hat{\phi^s}(\kb) 
\label{ppf1}
\ee
or using $\kb=\kbpe+k_{\parallel}\ep$\,:
\be
{\bf A}^s(\kb) = i {\bf k}\times \ep \, \hat{\psi^s}(\kb) - \kbpe 
k_{\parallel} \, \hat{\phi^s}(\kb) + \ep \, k^2_{\perp} \hat{\phi^s}(\kb) \, .
\label{ppf2}
\ee
The above equations indicate the relationships between the two orthogonal
systems (with ${\bf p}={\bf k}\times \ep$ and ${\bf q}={\bf k}\times {\bf p}$)
made of the triads $(\kb,{\bf p},{\bf q})$,
$(\ep, {\bf p}, \kbpe)$ and the system $(\ep, {\bf p}, \kb)$.
In terms of the decomposition used in 
Waleffe (1992)
%\cite{waleffe} 
with
\be
{\bf h}_{\pm}={\bf p}\times \kb \pm i {\bf p}
\label{wal}\ee
and writing ${\bf z}^s=A^s_+{\bf h}_+ + A^s_-{\bf h}_-$, it can be shown easily 
that $\psi^s=A^s_+-A^s_-$ and $\phi^s=A^s_++A^s_-$. In these latter variables, 
the s--energies $E^s$ are proportional to $\langle |A^s_+|^2+|A^s_-|^2\rangle$ 
and the s--helicities $\langle {\bf z}^s \cdot \nabla \times {\bf z}^s \rangle$
are proportional to $\langle |A^s_+|^2-|A^s_-|^2\rangle $. Note that $E^s$
is not a scalar\,: when going from a right-handed to a left-handed frame of 
reference, $E^s$ changes into $E^{-s}$.

%%%%%%%%%%%%%%%%%%%%%%%%%%%%%%%%%%%%%%%%%%%%%%%%%%%%%%%%%%%%%%%%%%%%%%%%%%%%%%%%
\subsection{Moments and cumulants}
%%%%%%%%%%%%%%%%%%%%%%%%%%%%%%%%%%%%%%%%%%%%%%%%%%%%%%%%%%%%%%%%%%%%%%%%%%%%%%%%

We now seek a closure for the energy tensor $q^{s\sp}_{j\jp}({\bf k})$ defined as
\be
\langle a^s_j(\kb) \, a^{\sp}_{\jp}(\kbp) \rangle
\equiv q^{s\sp}_{j\jp}(\kbp) \, \delta (\kb+ \kbp)
\label{ent}
\ee
in terms of second order moments of the two scalar fields $\hat{\psi^s}(\kb)$ 
and $\hat{\phi^s}(\kb)$. Simple manipulations lead, with the restriction $s=\sp$ 
(it can be shown that correlations with $\sp=-s$ have no long time influence 
and therefore are, for convenience of exposition, omitted), to\,: 
\be
\begin{array}{lll}
q_{11}^{ss}(\kbp)  &= \ k_2^2\Psi^s(\kb) -k_1k_2k_{\pa}I^s(\kb)
+k_{\pa}^2k_1^2 \Phi^s(\kb) \, , \\[.2cm]
q_{22}^{ss}(\kbp)  &= \ k_1^2\Psi^s(\kb) +k_1k_2k_{\pa}I^s(\kb)
+k_{\pa}^2k_2^2 \Phi^s(\kb)  \, , \\[.2cm]
q_{12}^{ss}(\kbp)+q_{21}^{ss}(\kbp) &= \ -2k_1k_2\Psi^s(\kb)
+k_{\pa}(k_1^2-k_2^2)I^s(\kb) +2k_1k_2k_{\pa}^2\Phi^s(\kb) \, , \\[.2cm]
q_{1\pa}^{ss}(\kbp)+q_{\pa 1}^{ss}(\kbp)  &= \ k_2 \kpe^2I^s(\kb)
-2k_1k_{\pa}\kpe^2\Phi^s(\kb) \, , \\[.2cm]
q_{2\pa}^{ss}(\kbp)+q_{\pa 2}^{ss}(\kbp)  &= \ -k_1 \kpe^2I^s(\kb)
-2k_2k_{\pa}\kpe^2\Phi^s(\kb) \, , \\[.2cm]
q_{\pa \ \pa}^{ss}(\kbp)  &= \ \kpe^4\Phi^s(\kb)  \, , \\[.2cm]
{{1}\over{k_1}}[q_{2\pa}^{ss}(\kbp)-q_{\pa2}^{ss}(\kbp)]  &= \
{{1}\over{k_2}}[q_{\pa1}^{ss}(\kbp)-q_{1\pa}^{ss}(\kbp)]   =
{{1}\over{k_{\pa}}}[q_{12}^{ss}(\kbp)-q_{21}^{ss}(\kbp)]  \\
& = -i\kpe^2R^s(\kb) \, ,
\end{array}
\label{corrq}
\ee
where the following correlators involving the toroidal and poloidal
fields have been introduced\,:
\be
\begin{array}{lll}
\langle \hat{\psi^s}(\kb) \hat{\psi^s}(\kbp) \rangle &=
\delta(\kb+\kbp)\Psi^s(\kbp) \, , \\[.2cm]
\langle \hat{\phi^s}(\kb) \hat{\phi^s}(\kbp) \rangle &=
\delta(\kb+\kbp)\Phi^s(\kbp) \, , \\[.2cm]
\langle \hat{\psi^s}(\kb) \hat{\phi^s}(\kbp) \rangle &=
\delta(\kb+\kbp)\Pi^s(-\kb) \, , \\[.2cm]
\langle \hat{\phi^s}(\kb) \hat{\psi^s}(\kbp) \rangle &=
\delta(\kb+\kbp)\Pi^s(\kb) \, , \\[.2cm]
R^s(\kb) &= \Pi^s(-\kb)+\Pi^s(\kb) \, , \\[.2cm]
I^s(\kb) &= i[\Pi^s(-\kb)-\Pi^s(\kb)] \, , \\
\end{array}
\label{corr}
\ee
and where $\kpe^2=k^2_1+k^2_2$\ , $k^2=\kpe^2+k^2_{\pa}$.
Note that $\Sigma_s R^s$ is the only
pseudo--scalar, linked to the lack of symmetry of the equations 
under plane reversal, \ie to a non--zero helicity.

The density energy spectrum writes
\be
e^s(\kb) = \Sigma_j \, q^{ss}_{jj} (\kb) = 
\kbpe^2(\Psi^s(\kb)+k^2\Phi^s(\kb)) \ .
\label{ess}
\ee
Note that it can be shown easily that the kinetic and magnetic energies
${{1}\over{2}} \langle u^2 \rangle$ and ${{1}\over{2}} \langle b^2 \rangle$
are equal in the context of the weak turbulence approximation.
Similarly, expressing the magnetic induction as a combination of ${\bf z}^{\pm}$
and thus of $\hat{\psi}^{\pm}$ and $\hat{\phi}^{\pm}$, the following symmetrized
cross--correlator of magnetic helicity (where the Alfv\'en velocity is used for
convenience) and its Fourier transform are found to 
be
\be
\oh \langle {\hat {\cal A}}_j({\kb}) \, {\hat b_j}({\kbp}) \rangle+
\oh \langle {\hat {\cal A}}_j({\kbp}) \, {\hat b_j}({\kb}) \rangle = \o4 \kbpe^2 
\Sigma_s R^s(\kb) \, \delta(\kb + \kbp) \, ,
\label{hm}\ee
where the correlations between the $+$ and $-$ variables are ignored because 
they are exponentially damped in the approximation of weak turbulence.
Similarly to the case of energy, there is equivalence between the kinetic and
magnetic helical variables in that approximation, hence the kinetic helicity 
defined as $\langle {\bf u} \cdot \pmbmath{\omega} \rangle$ writes simply in 
terms of its spectral density $H^V({\bf k})$\,:
\be
H^V(\kb) = k^2 H^M({\bf k}) = \o4 k^2 \kbpe^2 \Sigma_s R^s(\kb) \, .
\label{hv}\ee

In summary, the eight fundamental spectral density variables for which we seek 
a weak turbulence closure are the energy $e^s(\kb)$ of the three components 
of the ${\bf z}^s$ fields, the energy density along the direction of the
uniform magnetic field $\Phi^s(\kb)$, the correlators related to the 
off--diagonal terms of the spectral energy density tensor $I^s(\kb)$ and 
finally the only helicity--related pseudo--scalar correlators, 
namely $R^s(\kb)$. 

The main procedure that leads to a closure of weak turbulence for 
incompressible MHD is outlined in the Appendix. It leads to the equations 
(\ref{qss}) giving the temporal evolution of the
components of the spectral tensor $q^{s\sp}_{j\jp}(\kb)$ just defined.
The last technical step consists in transforming equations (\ref{qss}) of the 
Appendix in 
terms of the eight correlators we defined above. This leads us to the final set
of equations, constituting the kinetic equations for weak MHD turbulence.

%%%%%%%%%%%%%%%%%%%%%%%%%%%%%%%%%%%%%%%%%%%%%%%%%%%%%%%%%%%%%%%%%%%%%%%%%%%%%%%%
\subsection{The kinetic equations}
\label{keMHD}
%%%%%%%%%%%%%%%%%%%%%%%%%%%%%%%%%%%%%%%%%%%%%%%%%%%%%%%%%%%%%%%%%%%%%%%%%%%%%%%%

In the general case the kinetic equations for weak MHD turbulence are 

\begin{eqnarray}
\partial_t e^s ({\bf{k}}) = 
\label{K1}
\end{eqnarray}
$${\pi \varepsilon^2 \over b_0} \int \Bigg[ \left( L^2_{\bot} - 
\frac{X^2}{k^2}\right)\Psi^s ({\bf{L}}) - \left( k^2_{\bot} - 
\frac{X^2}{L^2}\right) \Psi^s ({\bf{k}}) + \left( L^2_{\bot}L^2 - 
\frac{k^2_{\pa} W^2}{k^2}\right)\Phi^s ({\bf{L}})$$
$$- \left(k^2_{\bot}k^2 - \frac{k^2_{\pa} Y^2}{L^2} \right)\Phi^s
({\bf{k}}) + \left( \frac{k_{\pa} XY}{L^2}\right)I^s ({\bf{k}})
- \left( \frac{k_{\pa} X W}{k^2}\right)I^s ({\bf{L}})\Bigg] 
Q^{-s}_k ({\pmbmath{\kappa}})$$
$$\delta(\kappa_{\pa}) \delta_{\kb, \kgb \Lb} \, d_{\kgb \Lb}$$

\begin{eqnarray}
\partial_t \left[k^2_{\bot} k^2 \Phi^s ({\bf{k}})\right] = 
\label{K2}
\end{eqnarray}
$${\pi \varepsilon^2 \over b_0} \int \Bigg[ k^2_{\pa} X^2 
\left(\frac{\Psi^s ({\bf{L}})}{k^2_{\bot} k^2} - \frac{\Phi^s 
({\bf{k}})}{L^2_{\bot}} \right) + \left(k_{\pa}^2 Z + k^2_{\bot} L^2_{\bot}
\right)^2 \left( \frac{\Phi^s({\bf{L}})}{k^2_{\bot} k^2} - 
\frac{\Phi^s({\bf{k}})}{L^2_{\bot} L^2} \right)$$
$$+ \frac{k_{\pa} X}{k^2_{\bot} k^2 }\left(k_{\pa}^2 Z + k^2_{\bot}L^2_{\bot}
\right)I^s ({\bf{L}}) + \left(\frac{k_{\pa} X Y}{2 L^2}\right)
I^s ({\bf{k}}) \Bigg] Q^{-s}_k ({\pmbmath{\kappa}}) \ \delta(\kappa_{\pa} ) \
\delta_{\kb, \kgb \Lb} \, d_{\kgb \Lb}$$
$$- \, {\varepsilon^2 \over b_0} s R^s ({\bf{k}}) {\cal P} \int \frac{X}{2 
\kappa_{\pa} L^2} \left(k_{\pa}Z - L_{\pa} k^2_{\bot} \right) Q^{-s}_k 
( {\pmbmath{\kappa}}) \ \delta_{\kb, \kgb \Lb} \ d_{\kgb \Lb}$$

\begin{eqnarray}
\partial_t \left[ k^2_{\bot} R^s ({\bf{k}})\right] = 
\label{K3}
\end{eqnarray}
$$ - {\pi \varepsilon \over b_0} \int \left[L^2_{\bot}
\left(\frac {Z + k^2_{\pa}}{k^2}\right) R^s({\bf{L}}) + \frac{k^2_{\bot}}{2}
\left(1 + \frac{(Z + k^2_{\pa})^2}{k^2 L^2}\right) R^s ({\bf{k}})
\right] Q^{-s}_k ({\pmbmath{\kappa}})$$
$$\delta (\kappa_{\pa} ) \ \delta_{\kb, \kgb \Lb} \, d_{\kgb \Lb} \, + \, 
{\varepsilon^2 \over b_0} s {\cal P} \int \Bigg[ 2 X \left(k_{\pa} Z - 
L_{\pa}k^2_{\bot} \right) \left(\Psi^s ({\bf{k}}) + k^2 \Phi^s({\bf{k}}) \right)$$
$$+ \left( {\left(k_{\pa} Z - L_{\pa} k^2_{\bot} \right)}^2 - k^2 X^2\right) I^s 
({\bf{k}}) \Bigg] \frac{Q^{-s}_{k} ({\pmbmath{\kappa}})}{2 \kappa_{\pa} k^2 L^2}
\ \delta_{\kb, \kgb \Lb} d_{\kgb \Lb}$$

\begin{eqnarray}
\partial_t \left[ k^2_{\bot} k^2 I^s ({\bf{k}})\right] = 
\label{K4}
\end{eqnarray}
$${\pi \varepsilon^2 \over b_0} \int \Bigg[ 
\left(L^2_{\bot}Z + \frac{k^2_{\pa}}{k^2_{\bot}}(Z^2 - X^2 )\right)I^s 
({\bf{L}}) + \left( \frac{k^2_{\pa} Y^2}{2 L^2} - k^2_{\bot} k^2 + 
\frac{k^2 X^2}{2 L^2} \right) I^s ({\bf{k}})$$
$$+ \left(\frac{k_{\pa} X Y}{L^2}\right) \Big( \Psi^s ({\bf{k}}) + 
k^2 \Phi^s ({\bf{k}}) \Big) + \frac{2k_{\pa} X}{k_{\bot}^2}
\Big( Z \Psi^s ({\bf{L}}) - \left( k^2_{\pa} Z + k^2_{\bot}L^2_{\bot} \right) 
\Phi^s ({\bf{L}}) \Big) \Bigg]$$ 
$$Q^{-s}_k ({\pmbmath{\kappa}}) \delta (\kappa_{\pa} ) \ \delta_{\kb, \kgb \Lb} 
\, d_{\kgb \Lb} \, - \, {\varepsilon^2 \over b_0} s R^s ({\bf{k}}) {\cal P} \int 
\frac{1} {2 \kappa_{\pa} L^2} \left( {\left( k_{\pa} Z - L_{\pa}k^2_{\bot}
\right)}^2 - k^2 X^2 \right)$$
$$Q^{-s}_k ({\pmbmath{\kappa}}) \ \delta_{\kb, \kgb \Lb} \ d_{\kgb \Lb}$$

\noindent with
$$\delta_{\kb, \kgb \Lb} = \delta({\bf{L}} + {\pmbmath{\kappa}} - {\bf{k}}), $$
$$d_{\kgb \Lb} = d\pmbmath{\kappa} \ d\bf{L},$$
and
\be
\begin{array}{lll}
Q^{-s}_k ({\pmbmath{\kappa}}) & = & k_m k_p \, q^{-s -s}_{p \ m}
({\pmbmath{\kappa}}) \\[.2cm]
&=&  X^2 \Psi^{-s} ({\pmbmath{\kappa}}) + 
X(k_{\pa} \kappa^2_{\bot} -\kappa_{\pa} Y)I^{-s} ({\pmbmath{\kappa}}) + 
(\kappa_{\pa} Y-k_{\pa} \kappa^2_{\bot})^2 \phi^{-s} ({\pmbmath{\kappa}}) \, .
\nonumber
\end{array}
\ee

Note that $Q^{-s}_k$ does not involve the spectral densities $R^s({\bf k})$,
because of symmetry properties of the equations. 
The geometrical coefficients appearing in the kinetic equations are
\begin{equation}
\begin{array}{lll}
X &= \ ({\bf{k}_{\bot}} \wedge {\pmbmath{\kappa}_{\bot}})_z \ = \  k_{\bot}
\kappa_{\bot}\sin\theta \, , \\[.2cm]
Y &= \ {\bf{k}_{\bot}} \cdot {\pmbmath{\kappa}_{\bot}} \ = \ k_{\bot}
\kappa_{\bot}\cos\theta \, , \\[.2cm]
Z &= \ {\bf{k}_{\bot}} \cdot {\bf{L}_{\bot}} \ = \ k^2_{\bot}- k_{\bot} 
\kappa_{\bot}\cos\theta = k^2_{\bot}-Y \, , \\[.2cm]
W &= \ {\bf{\pmbmath{\kappa}_{\bot}}} \cdot {\bf{L}_{\bot}} \ = 
\ k^2_{\bot}-L^2_{\bot} - k_{\bot}\kappa_{\bot}\cos\theta = Z-L^2_{\bot} \, , 
\end{array}
\label{}
\end{equation}
where $\theta$ is the angle between ${\bf k}_{\bot}$ and 
$\pmbmath{\kappa}_{\bot}$, and with 
\begin{equation}
d{\pmbmath{\kappa_{\bot}}} = \kappa_{\bot}d\kappa_{\bot}d\theta = 
\frac{L_{\bot}}{k_{\bot}\sin\theta} \ d\kappa_{\bot}dL_{\bot} \, ,
\label{}
\end{equation}
\begin{equation}
\cos\theta = 
\frac{\kappa^2_{\bot}+k^2_{\bot}-L^2_{\bot}}{2\kappa_{\bot}k_{\bot}} \, .
\label{}
\end{equation}
In (\ref{K2}), (\ref{K3}) and (\ref{K4}) ${\cal P} \int$ means the Cauchy 
Principal value of the integral in question.

%%%%%%%%%%%%%%%%%%%%%%%%%%%%%%%%%%%%%%%%%%%%%%%%%%%%%%%%%%%%%%%%%%%%%%%%%%%%%%%%
\section{General properties of the kinetic equations}
\label{pg}
\subsection{Dynamical decoupling in the direction parallel to ${\bf B_0}$}
%%%%%%%%%%%%%%%%%%%%%%%%%%%%%%%%%%%%%%%%%%%%%%%%%%%%%%%%%%%%%%%%%%%%%%%%%%%%%%%%

The integral on the right-hand side of the kinetic equation (\ref{K1}) contains 
a delta function of the form $\delta(\kappa_{\pa})$, the integration variable 
corresponding to the parallel component of one of the wavenumbers in the 
interacting triad. This delta function arises because of the three-wave 
frequency resonance condition. Thus, in any resonantly interacting wave triad 
(${\bf k},\ \kgb,\ {\bf L}$), there is always
one wave that corresponds to a purely 2D motion -- having no dependence on the
direction parallel to the uniform magnetic field -- whereas the other two 
waves have equal parallel components of their corresponding wavenumbers, \viz
$L_{\pa}=k_{\pa}$. This means that the parallel components of the wavenumber
enter in the kinetic equation of the total energy $e^s ({\bf{k}})$ as an external 
parameter and that the dynamics is decoupled at each level of $k_{\pa}$.
In other words, there is no transfer associated with the three-wave resonant 
interaction along the $k_{\pa}$-direction in ${\bf k}$-space for the total energy.
This result, using the exact kinetic equations developed here, corroborates what
has already been found in 
Montgomery and Turner (1981)
%\cite{montgo1} 
using a phenomenological analysis of 
the basic MHD equations, in Ng and Bhattacharjee (1996, 1997) in the 
framework of a model of weak MHD turbulence using individual wave packets, and
in Kinney and McWilliams (1998) with a Reduced MHD approach (RMHD). 

As for the kinetic equation (\ref{K1}), the other kinetic equations (\ref{K2}) 
to (\ref{K4}) have integrals containing delta functions of the form 
$\delta(\kappa_{\pa})$. But, in addition, they have PVIs which can, 
{\it a priori}, contribute to a transfer in the parallel direction. The eventual 
contributions of these PVIs are discussed in $\S \ref{noPVI}$.

%%%%%%%%%%%%%%%%%%%%%%%%%%%%%%%%%%%%%%%%%%%%%%%%%%%%%%%%%%%%%%%%%%%%%%%%%%%%%%%%
\subsection{Detailed energy conservation}
%%%%%%%%%%%%%%%%%%%%%%%%%%%%%%%%%%%%%%%%%%%%%%%%%%%%%%%%%%%%%%%%%%%%%%%%%%%%%%%%

Detailed conservation of energy for each interacting triad of waves
is a usual property in weak turbulence theory. 
This property is closely related with the frequency resonance condition
\[
\omega_{\bf k}=\omega_{\bf L}+ \omega_{\kgb},
\]
because $\omega$ can be interpreted as the energy of
one wave "quantum".
For Alfv\'en waves, the detailed energy conservation
 property is even stronger because one of the waves in any
resonant triad belongs to the 2D state with frequency equal to
zero,
\[
 \omega_{\kgb} \propto \kappa_{\pa} =0.
\]
Thus, for every triad of Alfv\'en waves ${\bf k}, {\bf L}$ and
${\kgb}$ (such that $ \kappa_{\pa} =0$) the energy is
conserved within two co-propagating waves having
wavevectors ${\bf k}$ and $ {\bf L}$. Mathematically, this corresponds
to the symmetry of the integrand in the equation for $e^s$ with
respect to changing  ${\bf k} \leftrightarrow {\bf L}$ (and correspondingly 
$\kgb = {\bf k} -  {\bf L} \to - \kgb$). 

As we have said, energy is conserved $k_{\pa}$ plane by $k_{\pa}$ plane so that, 
for each $k_{\pa}$, it can be shown from (\ref{K1}) 
\begin{equation}
{\partial \over \partial t} \int e^s (\kbpe,k_{\pa}) \, d\kbpe = 0 \, .
\label{econs}
\end{equation}

%%%%%%%%%%%%%%%%%%%%%%%%%%%%%%%%%%%%%%%%%%%%%%%%%%%%%%%%%%%%%%%%%%%%%%%%%%%%%%%%
\subsection{The magnetic and pseudo magnetic helicities}
%%%%%%%%%%%%%%%%%%%%%%%%%%%%%%%%%%%%%%%%%%%%%%%%%%%%%%%%%%%%%%%%%%%%%%%%%%%%%%%%

Since Woltjer (1958) we know that the magnetic helicity is an invariant of the 
MHD equations. However Stribling et al. (1994)
showed that in presence of a mean magnetic field ${\bf B_0}$ the part of the 
magnetic helicity associated with fluctuations is not conserved separately 
(whereas the total magnetic helicity, which takes into account a term 
proportional to ${\bf B_0}$, is, of course, an invariant). 
It is then interesting to know if, in the context of weak turbulence, the 
integral of the spectral density of fluctuations of the magnetic helicity is 
conserved, \ie
\begin{equation}
\int H^M({\bf k}) \, d\kb = \hbox{constant} \, ,
\label{hmt1}
\end{equation}
with
\begin{equation}
H^M({\bf k}) = \o4 k^2 \kbpe^2 \Sigma_s R^s(\kb) \, .
\label{hmt2}
\end{equation}
To investigate this point we define in the physical space the total magnetic 
helicity as 
\begin{equation}
H^M_T = \langle {\bf {\cal A_T}} \cdot {\bf b_T} \rangle \, ,
\label{hmt}
\end{equation}
where ${\bf b_T} = \nabla \times {\bf {\cal A_T}}$ and 
${\bf b_T} = {\bf b_0} + {\bf b}$. The magnetic induction equation 
\begin{equation}
\partial_t {\bf b_T} = \nabla \times ({\bf v} \times {\bf b_T})
\label{ind}
\end{equation}
implies that (Stribling et al. 1994) 
\begin{equation}
\partial_t H^M_T = \partial_t H^M + 2 {\bf b_0} \cdot \partial_t 
\langle {\bf {\cal A}} \rangle \, ,
\label{HMT}
\end{equation}
where $H^M$ is the magnetic helicity associated with fluctuations 
($H^M = \langle {\bf {\cal A}} \cdot {\bf b} \rangle$) and 
${\bf b}=\nabla \times {\bf {\cal A}}$. Direct numerical simulations (Stribling 
et al. 1994) show that the second term in the RHS of (\ref{HMT}) has a non-zero 
contribution to the total magnetic helicity, but in the context of weak 
turbulence the situation is different. Indeed, the magnetic induction equation 
leads also to the relation 
\begin{equation}
\partial_t \langle {\bf {\cal A}} \rangle = {1 \over 2} \langle {\bf z^-} 
\times {\bf z^+} \rangle \, .
\label{AT}
\end{equation}
Therefore the temporal evolution of the magnetic potential of fluctuations is
proportional to the cross product between z-fields of opposite polarities. As 
we have already pointed out, in the framework of weak turbulence this kind of 
correlation has no long time influence and thus the magnetic helicity associated 
with fluctuations appears to be an invariant of the weak turbulence equations. 
%Then it must be possible to check the relation (\ref{hmt1}) directly from the 
%kinetic equations. Unfortunately our attempts to show 
%this property analytically failed because of 
%the complexity of the full 3D kinetic equations. 
We leave for the future the 
investigation of this point\,: in particular it will be helpful to make 
numerical computations to show, at least at this level, the invariance of the
magnetic helicity. 

The correlators $R^s ({\bf{k}})$ and $I^s ({\bf{k})}$ have been defined in the 
previous section as the real part and the imaginary part of $\Pi^s(\kb)$, the
cross--correlator of the toroidal field $\hat{\psi^s}(\kb)$ and of the poloidal
field $\hat{\phi^s}(\kb)$. Then $\kbpe^2 \Sigma_s R^s(\kb)$ appears as the 
spectral density of the magnetic helicity. On the other hand $I^s ({\bf{k})}$, 
which we will call the anisotropy correlator (or pseudo magnetic helicity), is 
neither a conserved quantity nor a positive definite quantity. 
Although $R^s$ and $I^s$ evolve according to their own kinetic equations 
(\ref{K4}) and (\ref{K3}), the range of values they can take on is bounded by 
$\Psi^s$ and  $\Phi^s$, with the bounds being a simple consequence of
the definition of these quantities. Two realizability conditions (see also 
Cambon and Jacquin 1989; Cambon et al. 1997) between the four
correlators $\Psi^s$, $\Phi^s$, $I^s$ and $R^s$ can be obtained  from
\begin{equation}
\langle \, |\hat{\psi^s}(\kb) \pm k \hat{\phi^s}(\kb)|^2 \, \rangle \ge 0 \, ,
\end{equation}
and
\begin{equation}
\langle \, |\hat{\psi^s}(\kb)|^2 \, \rangle \langle \, |\hat{\phi^s}(\kb)|^2 
\, \rangle \ge | \langle \,\hat{\psi^s}(\kb) \hat{\phi^s}(-\kb) \, 
\rangle |^2 \, .
\end{equation}
These conditions are found to be respectively 
\begin{equation}
\Psi^s(\kb) + k^2 \Phi^s(\kb) \ge | k R^s(\kb)| \, ,
\label{bound1}
\end{equation}
and
\begin{equation}
4 \Psi^s(\kb) \Phi^s(\kb) \ge {R^s}^2(\kb) + {I^s}^2(\kb) \, .
\label{bound}
\end{equation}
Note that the combination
\begin{equation}
{\cal Z}=(1/2) k_{\perp}^2 [k^2 \Phi({\bf{k}}) - \Psi({\bf{k}}) - 
i|k|I({\bf{k}})]
\end{equation}
is named polarization anisotropy in 
Cambon and Jacquin (1989).
%\cite{cambon1}. 
The consequences of the
realizability conditions is explained below.

%%%%%%%%%%%%%%%%%%%%%%%%%%%%%%%%%%%%%%%%%%%%%%%%%%%%%%%%%%%%%%%%%%%%%%%%%%%%%%%%
\subsection{Purely 2D modes and two-dimensionalisation of 3D spectra}
\label{noPVI}
%%%%%%%%%%%%%%%%%%%%%%%%%%%%%%%%%%%%%%%%%%%%%%%%%%%%%%%%%%%%%%%%%%%%%%%%%%%%%%%%

The first consequence of the fact that there is no transfer of the total energy
in the $k_{\pa}$ direction in ${\bf k}$-space is an asymptotic 
two-dimensionalisation of the energy spectrum $e^s ({\bf{k}})$. Namely, the 3D 
initial spectrum spreads over the transverse wavenumbers, ${\bf k}_\perp$, but 
remains of the same size in the $k_{\pa}$ direction, and the support of the 
spectrum becomes very flat (pancake-like) for large time.
The two-dimensionalisation of weak MHD turbulence has been observed in laboratory 
experiments (Robinson and Rusbridge 1971), in the solar wind data (Bavassano et 
al. 1982; Matthaeus et al. 1990; Horbury et al. 1995; Bieber et al. 1996), 
and in many direct numerical simulations of the three--dimensional MHD equations 
(Oughton et al. 1994) or of the RMHD equations (Kinney and McWilliams 1998). 

From a mathematical point of view, the two-dimensionalisation of the total 
energy means that, for large time, the energy spectrum $e^s ({\bf{k}})$ is 
supported on a volume of wavenumbers such that for most of them 
${ k}_\perp \gg k_{\pa}$. This implies that
$\Psi^s({\bf{k}})$ and $\Phi^s({\bf{k}})$ 
are also supported on the same anisotropic region of wavenumbers because both of
them are non-negative.
This, in turn, implies that both $R^s$ and $I^s$ will also be non-zero only for 
the same region in the ${\bf k}$-space as $e^s ({\bf{k}})$, $\Psi^s({\bf{k}})$
and $\Phi^s({\bf{k}})$, as it follows from the bound (\ref{bound1}) and 
(\ref{bound}). This fact allows one to expand the integrands in the kinetic 
equations in powers of small $k_{\pa}/k_{\perp}$. 
At the leading order in $k_{\pa}/k_{\perp}$, one obtains
\be
\begin{array}{lll}
\hspace{-.3cm}
\partial_t \left[k^2_{\bot} \Psi^s ({\bf{k}})\right] = 
\end{array}
\label{K5}
\ee
$${\pi \varepsilon^2 
\over b_0} \int \Bigg[ \left( L^2_{\bot} - \frac{X^2}{k_{\bot}^2}\right)\Psi^s 
({\bf{L}}) - \left( k^2_{\bot} - \frac{X^2}{L_{\bot}^2}\right) \Psi^s({\bf{k}})
\Bigg] X^2 \Psi^{-s} ({\pmbmath{\kappa}}) \ \delta(\kappa_{\pa}) \ 
\delta_{\kb, \kgb \Lb} \, d_{\kgb \Lb} \, , $$

\be
\begin{array}{lll}
\hspace{-.3cm}
\partial_t \left[k^4_{\bot} \Phi^s ({\bf{k}})\right] = 
\end{array}
\label{K6} 
\ee
$${\pi \varepsilon^2 \over b_0} \int \Bigg[ L^4_{\bot} \Phi^s({\bf{L}}) 
- k^4_{\bot} \Phi^s({\bf{k}}) \Bigg] X^2 \Psi^{-s} ({\pmbmath{\kappa}})
\ \delta(\kappa_{\pa} ) \ \delta_{\kb, \kgb \Lb} \, d_{\kgb \Lb} \, ,$$

\be
\begin{array}{lll}
\hspace{-.3cm}
\partial_t \left[ k^2_{\bot}R^s ({\bf{k}})\right] = 
\end{array}
\label{K7} 
\ee
$$- {\pi \varepsilon^2 \over 
b_0}\int \left[ \left(\frac {L^2_{\bot} Z}{k_{\bot}^2}\right) 
R^s({\bf{L}}) + \left(\frac{k^2_{\bot}}{2} + \frac{Z^2}{2 L_{\bot}^2}\right) 
R^s ({\bf{k}}) \right] X^2 \Psi^{-s} ({\pmbmath{\kappa}}) \ \delta 
(\kappa_{\pa} ) \ \delta_{\kb, \kgb \Lb} \, d_{\kgb \Lb} \, , $$

\be
\begin{array}{lll}
\hspace{-.3cm}
\partial_t \left[ k^2_{\bot} I^s ({\bf{k}})\right] = 
\end{array}
\label{K8} 
\ee
$${\pi \varepsilon^2 \over 
b_0}\int \Bigg[ \left(\frac {L^2_{\bot} Z}{k_{\bot}^2}\right)
I^s ({\bf{L}}) - \left( k^2_{\bot} - \frac{X^2}{2 L_{\bot}^2} \right) 
I^s ({\bf{k}}) \Bigg] X^2 \Psi^{-s} ({\pmbmath{\kappa}}) \ \delta(\kappa_{\pa} )
\ \delta_{\kb, \kgb \Lb} \, d_{\kgb \Lb} \ . $$
Note that the principal value terms drop out of the kinetic equations at leading
order. This property means that there is no transfer of any of the eight 
correlators in the $k_{\pa}$ direction in ${\bf k}$-space. 

One can see from the above that the equations for the toroidal and poloidal
energies decouple for large time. These equations describe the shear-Alfv\'en and 
pseudo-Alfv\'en waves respectively. An energy exchange between $\Psi^s ({\bf{k}})$
and $\Phi^s ({\bf{k}})$ is however possible in an initial phase, \ie before the 
two-dimensionalisation of the spectra. A preliminary investigation 
shows that this exchange is actually essentially generated by the 
magnetic helicity through the principal value terms\,: the magnetic helicity 
plays the role of a catalyst which transfers toroidal energy into poloidal energy. 
On the other hand, in the large time limit, the magnetic helicity $\Sigma_s R^s$
and the pseudo magnetic helicity $I^s$ are also described by equations which 
are decoupled from each other and from the toroidal and poloidal energies. 
It is interesting that the kinetic equation for the shear-Alfv\'en waves
(\ie for $\Psi^s ({\bf{k}})$) can be obtained also from the RMHD 
equations which have been derived under the same conditions of quasi
two-dimensionality (see \eg Strauss 1976). 
%However, the kinetic equations for the pseudo-Alfv\'en waves and for the
%helicities cannot be obtained from the RMHD ansatz which does not consider
%these quantities.

An important consequence of the dynamical decoupling at different $k_{\pa}$'s 
within the kinetic equation formalism
is that the set of purely 2D modes (corresponding to $k_{\pa}=0$) evolve 
independently of the 3D part of the spectrum (with $k_{\pa} \ne 0$) and can be 
studied separately. One can interpret this fact as a neutral stability of the
purely 2D state with respect to 3D perturbations. As we mentioned in the 
Introduction, the kinetic equations themselves are applicable
to a description of  $k_{\pa} = 0$ modes only if the correlations of the dynamical
fields decay in all directions, so that their spectra are sufficiently smooth
for all wavenumbers including the ones with  $k_{\pa} = 0$. To be precise,
the characteristic  $k_{\pa}$ over which the spectra can experience 
significant changes must be greater than $\epsilon^2$.
Study of such 2D limits of 3D spectrum will be presented in the next section.
It is possible, however, that in some physical situations the correlations
decay slowly along the magnetic field
due to a (hypothetical) energy  condensation at the $k_{\pa} = 0$ modes.
In this case, the modes with $k_{\pa} = 0$ should be treated as a separate
component, a condensate, which modifies the dynamics of the 3D modes
in a manner somewhat similar to the superfluid condensate, as described by 
Bogoliubov (Landau and Lifshitz 1968). We leave this problem for future study.

%%%%%%%%%%%%%%%%%%%%%%%%%%%%%%%%%%%%%%%%%%%%%%%%%%%%%%%%%%%%%%%%%%%%%%%%%%%%%%%%
\subsection{Asymptotic solution of the 3D kinetic equations}
%%%%%%%%%%%%%%%%%%%%%%%%%%%%%%%%%%%%%%%%%%%%%%%%%%%%%%%%%%%%%%%%%%%%%%%%%%%%%%%%

The parallel wavenumber $k_\pa$ enters equations (\ref{K5})-(\ref{K8}) 
only as an external parameter. In other words,
the wavenumber space is foliated into the dynamically 
decoupled planes  $k_\pa=0$. Thus, the large-time asymptotic solution
can be found in the following form,

\begin{eqnarray}
\Psi^s ({\bf{k}_\perp}, k_\pa) &=&  f_1 (k_\pa) \Psi^s ({\bf{k}}_\perp,0), \\
\Phi^s ({\bf{k}_\perp}, k_\pa) &=&  f_2 (k_\pa) \Phi^s ({\bf{k}}_\perp,0), \\
R^s ({\bf{k}_\perp}, k_\pa) &=&  f_3 (k_\pa) R^s ({\bf{k}}_\perp,0), \\
I^s ({\bf{k}_\perp}, k_\pa) &=&  f_4 (k_\pa) I^s ({\bf{k}}_\perp,0), 
\end{eqnarray}
where $f_i, \; (i=1,2,3,4) $ are some arbitrary functions of 
$k_\pa$ satisfying the conditions $f_i(0)=1$ (and such that the
bounds  (\ref{bound1}) and (\ref{bound}) are satisfied).
Substituting these formulae into (\ref{K5})-(\ref{K8}), one can readily see
that the functions  $f_i$ drop out of the problem, and the 
solution of the 3D equations is reduced to solving a 2D
problem for $\Psi^s ({\bf{k}}_\perp,0), \Phi^s ({\bf{k}}_\perp,0), 
R^s ({\bf{k}}_\perp,0)$ and $ I^s ({\bf{k}}_\perp,0)$, which
will be described in the next section.

%%%%%%%%%%%%%%%%%%%%%%%%%%%%%%%%%%%%%%%%%%%%%%%%%%%%%%%%%%%%%%%%%%%%%%%%%%%%%%%%
\section{Two--dimensional problem}
\label{swt4}
%%%%%%%%%%%%%%%%%%%%%%%%%%%%%%%%%%%%%%%%%%%%%%%%%%%%%%%%%%%%%%%%%%%%%%%%%%%%%%%%

Let us consider Alfv\'en wave turbulence which is axially symmetric with respect 
to the external magnetic field. Then $ I^s ({{k}}_\perp,0)=0$ because
of the condition  $ I^s (- {\bf{k}})= - I^s ({\bf{k}})$. In the following,
we will consider only solutions with $R^s=0$. (One can easily
see that $R^s$ will remain zero if it is zero initially.)
The remaining equations to be solved are
\begin{eqnarray}
\frac{\partial E^s_{\bot} (k_{\bot},0)}{\partial t} = 
\label{equ2D_1}
\end{eqnarray}
$${\pi \varepsilon^2 \over b_0} \int (\eh_L \cdot \eh_k)^2 \, sin \theta \ 
\frac{k_{\bot}}{\kappa_{\bot}}
\ E^{-s}_{\bot} (\kappa_{\bot},0) \ \Bigg[ k_{\bot} 
E^s_{\bot} (L_{\bot},0) - L_{\bot} E^s_{\bot} (k_{\bot},0) \Bigg] 
d{\kappa_{\bot}} d{L_{\bot}} \, , $$
\begin{eqnarray}
\frac{\partial E^s_{\parallel} (k_{\bot},0)}{\partial t} = 
\label{equ2D_2}
\end{eqnarray}
$${\pi \varepsilon^2 \over b_0} \int sin \theta \ \frac{k_{\bot}}{\kappa_{\bot}}
\ E^{-s}_{\bot} (\kappa_{\bot},0) \ \Bigg[ k_{\bot} 
E^s_{\parallel} (L_{\bot},0) - L_{\bot} E^s_{\parallel} (k_{\bot},0) \Bigg] 
d{\kappa_{\bot}} d{L_{\bot}} \, , $$
where $\eh_k$ and $\eh_L$ are the unit vectors along $\bf k_{\bot}$ and 
$\bf L_{\bot}$ respectively\, and
\begin{equation}
E^s_{\bot}(k_{\bot},0) = k^3_{\bot} \Psi^s(k_{\bot},0) \ ,
\end{equation}
\begin{equation}
E^s_{\pa}(k_{\bot},0) = k^5_{\bot} \Phi^s(k_{\bot},0) \ ,
\end{equation}
are the horizontal and the vertical components of the energy density. Thus, we 
reduced the original 3D problem to finding solution for the purely 2D state. It
may seem unusual that strongly turbulent 2D vortices (no waves for $k_{\pa}=0$\,!) 
are described by the kinetic equations obtained for weakly turbulent waves. 
Implicitly, this fact relies on continuity of the 3D spectra near $k_{\pa}=0$, so
that one could take the limit $k_{\pa} \to 0$. In real physical situations such 
continuity results from the fact that the external  magnetic field is not perfectly
unidirectional and, therefore, there is a natural smoothing of the spectrum over
a small range of angles. 

The equation (\ref{equ2D_1}) corresponds to
the evolution of the  shear-Alfv\'en waves for which the energy 
fluctuations are transverse to ${\bf B_0}$ whereas equation (\ref{equ2D_2})
describes  the pseudo-Alfv\'en waves for which the fluctuations 
are along ${\bf B_0}$. Both waves propagate along ${\bf B_0}$ at the same 
Alfv\'en speed. Equation (\ref{equ2D_1}) describes the interaction 
between two shear-Alfv\'en waves, $E^{\pm}_{\bot}$, propagating 
in opposite directions. On the other hand, the evolution of the pseudo-Alfv\'en 
waves depend on their interactions with the shear-Alfv\'en waves. The detailed 
energy conservation of the equation (\ref{equ2D_1}) implies that there is no
exchange of energy between the two different kinds of waves. 
The physical picture in this case is that the shear-Alfv\'en 
waves interact only among themselves and 
evolve independently of the pseudo-Alfv\'en waves. The pseudo-Alfv\'en waves 
scatter from the shear-Alfv\'en 
waves without amplification or damping and they do not interact with each other.

Using a standard two--point closure of turbulence (see \eg Lesieur 1990)
in which the characteristic 
time of transfer of energy is assumed known and written {\it a priori},
namely the EDQNM closure, 
Goldreich and Sridhar (1995)
%\cite{gosri} 
derived a variant 
of the kinetic equation (\ref{equ2D_1}) but for strong anisotropic
MHD turbulence.
In their analysis, the ensuing energy spectrum, which 
depends (as it is well known) on the phenomenological evaluation of the 
characteristic transfer time, thus differs from our result where the 
dynamics is self--consistent, closure being obtained through the 
assumption of weak turbulence.

It can be easily verified that the geometrical coefficient appearing in the
closure equation in 
Goldreich and Sridhar (1995)
%\cite{gosri} 
is identical to the one we find for the
$E^s_{\bot}({\bf k}_{\bot},k_{\pa})$ spectrum in the two--dimensional case.
However, the two formulations, beyond the above discussion on
characteristic time scales, differ in a number of ways\,:
(i) We choose to let the flow variables to be non mirror--symmetric, whereas
helicity is not taken into account in 
Goldreich and Sridhar (1995)
%\cite{gosri} 
where they have
implicitly assumed $R^s\equiv 0$; 
(ii) However, because of the anisotropy introduced by the presence of a
uniform magnetic field, one must take into account the coupled dynamics
of the energy of the shear Alfv\'en waves, the pseudo--Alfv\'en wave and the
pseudo magnetic helicity $I^s$\,; indeed, even if initially $I^s\equiv 0$, it is
produced by wave coupling and is part of the dynamics.
(iii) In three dimensions, all geometrical coefficients that depend on
$k^2=k^2_{\bot}+k^2_{\pa}$ have a $k_{\pa}$--dependence which is a function
of initial conditions and again is part of the dynamics.

%%%%%%%%%%%%%%%%%%%%%%%%%%%%%%%%%%%%%%%%%%%%%%%%%%%%%%%%%%%%%%%%%%%%%%%%%%%%%%%%
\subsection{Kolmogorov spectra}
\subsubsection{The Zakharov transformation}
%%%%%%%%%%%%%%%%%%%%%%%%%%%%%%%%%%%%%%%%%%%%%%%%%%%%%%%%%%%%%%%%%%%%%%%%%%%%%%%%

\begin{figure}
\centerline{\psfig{file=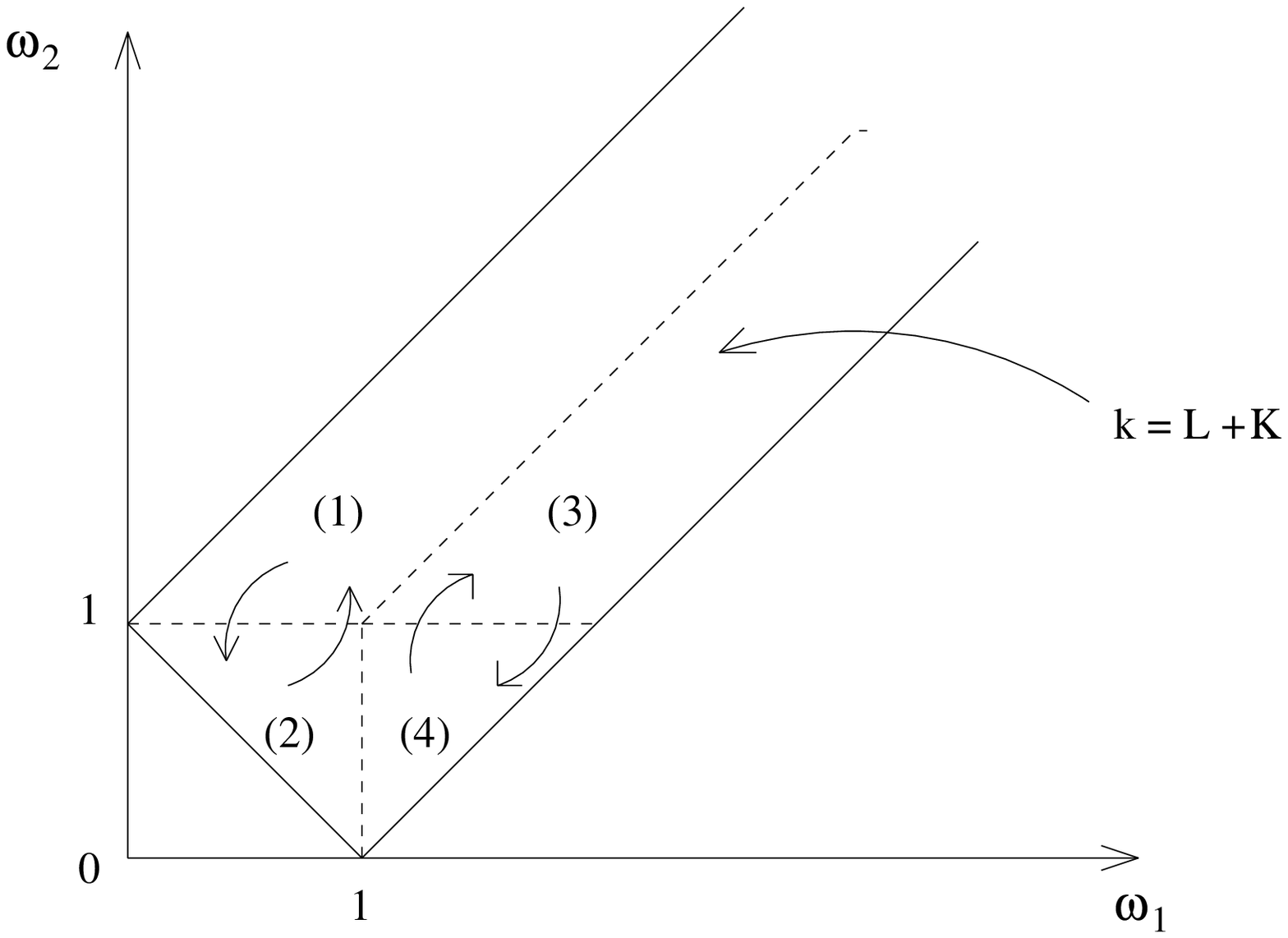,width=10cm,clip=}}
\caption[]{Geometrical representation of the Zakharov transformation. The 
rectangular region, corresponding to the triad interaction $\kbpe = \Lb + \kgb$, 
is decomposed into four different regions called (1), (2), (3) and (4); 
$\omega_1$ and $\omega_2$ are respectively the dimensionless variables 
$\kappa_{\bot} / k_{\bot}$ and $L_{\bot} / k_{\bot}$. The Zakharov transformation 
applied to the collision integral consists in exchanging regions (1) and (2), 
and regions (3) and (4).}
\label{fzakh}
\end{figure}

The symmetry of the previous equations allows us to perform a conformal 
transformation, called the Zakharov transformation (also used in modeling of 
strong turbulence, see Kraichnan 1967), in order to find the exact stationary 
solutions of the kinetic equations as power laws (Zakharov et al. 1992). This 
operation (see Figure \ref{fzakh}) consists of writing the kinetic equations in 
dimensionless variables $\omega_1 = \kappa_{\bot} / k_{\bot}$ and 
$\omega_2 = L_{\bot} / k_{\bot}$, setting $E^{\pm}_{\bot}$ by 
$k_{\bot}^{n_{\pm}}$, and then rearranging the collision integral by the 
transformation 
\begin{equation}
\omega_1^{\prime} = \frac{\omega_1}{\omega_2} \ ,
\end{equation}
and
\begin{equation}
\omega_2^{\prime} = \frac{1}{\omega_2} \ . 
\end{equation}
The new form of the collision integral, resulting from the summation of the 
integrand in its primary form and after the Zakharov transformation, is 
$$\partial_t E^s_{\bot} \sim 
\int \left( {\omega_2^2+1-\omega_1^2 \over 2\omega_2} \right)^2 
\left( 1 - ({\omega_1^2+1-\omega_2^2 \over 2 \omega_1})^2 \right)^{1/2}
\omega_1^{n_{-s}-1} \omega_2$$
$$( \omega_2^{n_s-1} -1 ) 
(1 - \omega_2^{-n_s-n_{-s}-4}) \, d\omega_1 \, d\omega_2 \, . 
$$
The collision integral can be null for specific values of $n_{\pm}$. 
The exact solutions, called the Kolmogorov spectra, correspond to these values 
which satisfy 
\begin{equation}
n_+ + n_- = -4 \ .
\end{equation}
%\begin{figure}
%\centerline{\psfig{file=Fig1_wt.ps,width=10cm,clip=}}
%\caption[]{Geometrical representation of the Zakharov transformation. The 
%rectangular region, corresponding to the triad interaction $\kbpe = \Lb + \kgb$, 
%is decomposed into four different regions called (1), (2), (3) and (4); 
%$\omega_1$ and $\omega_2$ are respectively the dimensionless variables 
%$\kappa_{\bot} / k_{\bot}$ and $L_{\bot} / k_{\bot}$. The Zakharov transformation 
%applied to the collision integral consists in exchanging regions (1) and (2), 
%and regions (3) and (4).}
%\label{fzakh}
%\end{figure}
It is important to understand that the Zakharov transformation is not an identity
transformation, and it can lead to spurious solutions. The necessary and 
sufficient condition for a spectrum obtained by the Zakharov transformation to be 
a solution of the kinetic equation is that the right hand side integral in 
(\ref{equ2D_1}) (\ie before the Zakharov transformation) equation converges. 
This condition is called the locality of the spectrum and leads to the following 
restriction on the spectral indices in our case\,: 
\begin{equation}
-3 < n_{\pm} < -1 \ .
\label{locality}
\end{equation}
A detailed study of the Kolmogorov spectrum locality will be given in section
\ref{var}.

In the particular case of a zero cross--correlation one has $E^+_{\bot} = 
E^-_{\bot} = E_{\bot} \sim k^n_{\bot}$ with only one solution 
$$n=-2\ .$$
Note that the thermodynamic equilibrium, corresponding to the equipartition 
state for which the flux of energy is zero instead of being finite as in the
above spectral forms, corresponds to the choices $n_+=n_-=1$ for both the 
perpendicular and the parallel components of the energy.

%%%%%%%%%%%%%%%%%%%%%%%%%%%%%%%%%%%%%%%%%%%%%%%%%%%%%%%%%%%%%%%%%%%%%%%%%%%%%%%%
\subsubsection{The Kolmogorov constants $C_K(n_s)$ and $C^{\prime}_K(n_s)$}
%%%%%%%%%%%%%%%%%%%%%%%%%%%%%%%%%%%%%%%%%%%%%%%%%%%%%%%%%%%%%%%%%%%%%%%%%%%%%%%%

The final expression of the Kolmogorov--like spectra found above
as a function of the Kolmogorov constant (generalised to MHD)
$C_K(n_s)$ and of the flux of energy $P^s_{\bot}(k_{\bot})$ can be 
obtained in the following way. For a better understanding, the demonstration 
will be done in the simplified case of a zero cross--correlation. The 
generalization to the correlated case ($E^+\not= E^-$) is straightforward.
Using the definition of the flux, 
\begin{equation}
\partial_t E_{\bot} (k_{\bot},0) = - \partial_{k_{\bot}}
P_{\bot}(k_{\bot}) \ ,
\end{equation}
one can write the flux of energy as a function of the collision integral (with the 
new form of the integrand) depending on $n$. Then the limit $n \to -2$ is 
taken in order to have a constant flux $P_{\bot}$ with no more dependence in 
$k_{\bot}$, as it is expected for a stationary spectrum in the inertial range. 
Here we have considered an infinite inertial range to use the Zakharov 
transformation. 
Whereas the collision integral tends to zero when $n \to -2$, the limit with 
which we are concerned is not zero because of the presence of a denominator 
proportional to $2n+4$, and which is a signature of the dimension in 
wavenumber of the flux. Finally the ``L'Hospital's rule'' gives the value of 
$P_{\bot}$ from which it is possible to write the Kolmogorov spectrum of the 
shear-Alfv\'en waves 
\begin{equation}
E_{\bot}(k_{\bot},0) = P^{1/2}_{\bot} \ C_K(-2) \ k^{-2}_{\bot} \ ,
\end{equation}
with the Kolmogorov constant 
\begin{equation}
C_K(n) = \sqrt{\frac{-2 b_0}{\pi \epsilon^2 J_1(n)}} \ ,
\end{equation}
and with the following form for the integral $J_1(n)$ 
\begin{equation}
J_1(n) = 2^{n+3} \int_{x=1}^{+ \infty} \int_{y=-1}^1 
\frac{\sqrt{(x^2-1)(1-y^2)} \ (xy+1)^2}{(x-y)^{n+6} \ (x+y)^{2-n}} 
\end{equation}
$$\left[2^{1-n} - (x+y)^{1-n}\right] \ln \left(\frac{x+y}{2}\right)\,dxdy \ .$$
As expected, the calculation gives a negative value for the integral $J_1(n)$ and 
for the particular value $n=-2$, we obtain $C_K(-2) \simeq 0.585$. Note that the 
integral $J_1(n)$ converges only for $-3 < n < -1$. 

\begin{figure}
\centerline{\psfig{file=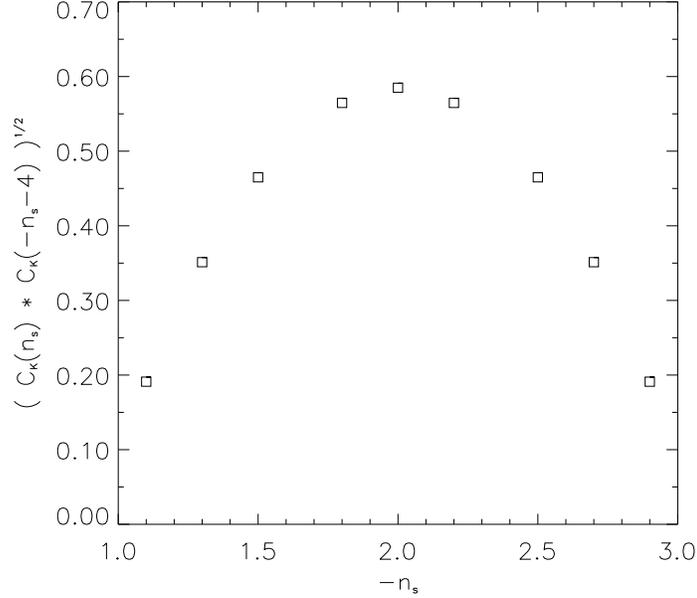,width=10cm,clip=}}
\caption[]{Variation of $\sqrt{C_K(n_s)\,C_K(-n_s-4)}$ as a function of $-n_s$. 
Notice the symmetry around the value $-n_s$ corresponding to the case of zero 
velocity-magnetic field correlation.}
\label{f1}
\end{figure}

\begin{figure}
\centerline{\psfig{file=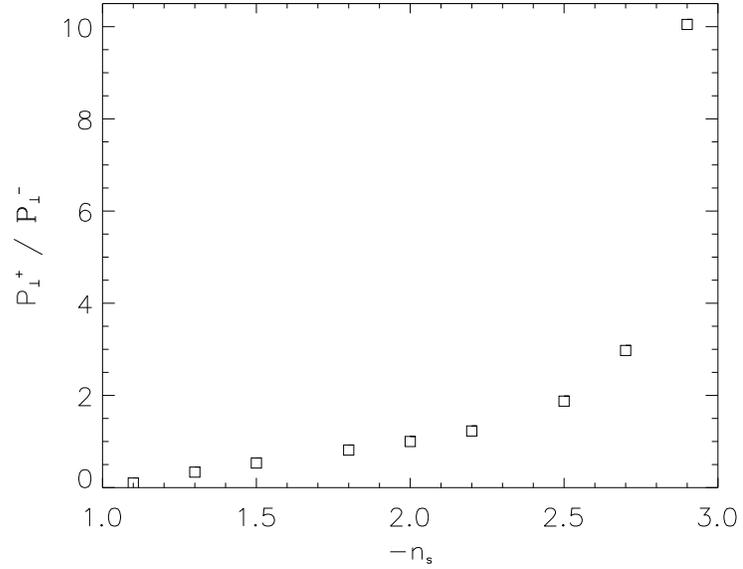,width=11cm,clip=}}
\caption[]{Variation of $P^+_{\bot} / P^-_{\bot}$, the ratio of fluxes of energy, 
as a function of $-n_s$. For the zero cross correlation case the ratio is $1$.}
\label{f2}
\end{figure}

The generalization to the case of non--zero cross--correlation gives 
the relations
\begin{eqnarray}
E^+_{\bot}(k_{\bot},0) \ E^-_{\bot}(k_{\bot},0) &=&
P^+_{\bot} \ C^2_K(n_s) \ k^{-4}_{\bot} = P^-_{\bot} \ 
C^2_K(-n_s-4) \ k^{-4}_{\bot} \nonumber \\
&=& \sqrt{P^+_{\bot} P^-_{\bot}} \, C_K(n_s) \, 
C_K(-n_s-4) \, k^{-4}_{\bot} \ ,
\label{pek4}
\end{eqnarray}
where the second formulation is useful to show the symmetry with respect to $s$.
The computation of the Kolmogorov constant $C_K$ as a function of $-n_s$ is given 
in Figure \ref{f1}. An asymmetric form is observed which means that the ratio 
$P^+_{\bot} / P^-_{\bot}$ is not constant, as we can see in Figure \ref{f2} 
where we plot this ratio as a function of $-n_s$. We see that for any ratio 
$P^+_{\bot} / P^-_{\bot}$ there corresponds a unique value of $n_s$, between the 
singular ratios $P^+_{\bot} / P^-_{\bot} = + \infty$ for $n_s=-3$ and 
$P^+_{\bot} / P^-_{\bot} = 0$ for $n_s=-1$. Thus, a larger flux of energy $P^+$ 
corresponds to a steeper slope of the energy spectra $E^+_{\bot}(k_{\bot},0)$
in agreement with the physical image that a larger flux of energy implies a 
faster energy cascade.

%\begin{figure}
%\centerline{\psfig{file=Fig2_wt.ps,width=10cm,clip=}}
%\caption[]{Variation of $\sqrt{C_K(n_s)\,C_K(-n_s-4)}$ as a function of $-n_s$. 
%Notice the symmetry around the value $-n_s=2$ corresponding to the case of zero 
%velocity-magnetic field correlation.}
%\label{f1}
%\end{figure}
%\begin{figure}
%\centerline{\psfig{file=Fig3_wt.ps,width=11cm,clip=}}
%\caption[]{Variation of $P^+_{\bot} / P^-_{\bot}$, the ratio of fluxes of energy, 
%as a function of $-n_s$. For the zero cross correlation case the ratio is $1$.}
%\label{f2}
%\end{figure}
%\begin{figure}
%\centerline{\psfig{file=Fig4_wt.ps,width=11cm,clip=}}
%\caption[]{Variation of $\sqrt{C^{\prime}_K(n_s) \ C^{\prime}_K(-n_s-4)}$ as a 
%function of $-n_s$. Notice the symmetry around the value $-n_s=2$ corresponding to 
%the zero cross correlation case.}
%\label{f3}
%\end{figure}

In the zero cross--correlation case, a similar demonstration for the 
pseudo-Alfv\'en waves $E^s_{\parallel}(k_{\bot},0)$ leads to the relation
\begin{equation}
E_{\parallel}(k_{\bot},0) = P_{\parallel} P^{-1/2}_{\bot} \ 
C^{\prime}_K (-2) \ k^{-2}_{\bot} \ ,
\end{equation}
with the general form of the Kolmogorov constant
\begin{equation}
C^{\prime}_K(n) = \sqrt{\frac{-2 b_0 J_1(n)}{\pi \epsilon^2 J_2(n) J_2(-n-4)}} \ ,
\end{equation}
where the integral $J_2(n)$ is
\begin{equation}
J_2(n) = 2^{n+3} \int_{x=1}^{+ \infty} \int_{y=-1}^1 
\frac{\sqrt{(x^2-1)(1-y^2)}}{(x-y)^{n+6} \ (x+y)^{2-n}} 
\end{equation}
$$\left[2^{1-n} - (x+y)^{1-n}\right] \ln \left(\frac{x+y}{2}\right)\,dxdy \ .$$

\noindent Note that the integral $J_2(n)$ converges only for $-3 < n < -1$. 
The presence of the flux $P_{\bot}$ in the Kolmogorov spectrum is 
linked to the presence of $E_{\bot}$ in the kinetic equation of $E_{\parallel}$.
A numerical evaluation gives $C^{\prime}_K(-2) \simeq 0.0675$ whereas the 
generalization for the non--zero cross--correlation is 
\begin{eqnarray}
E^+_{\parallel}(k_{\bot},0) \ E^-_{\parallel}(k_{\bot},0) &=&
\frac{P^+_{\parallel} P^-_{\parallel}}{P^+_{\bot}} \ C^{\prime 2}_K(n_s) 
\ k^{-4}_{\bot} = 
\frac{P^+_{\parallel} P^-_{\parallel}} {P^-_{\bot}} \ C^{\prime 2}_K(-n_s-4) 
\ k^{-4}_{\bot} \nonumber\\
&=& \frac{P^+_{\parallel} P^-_{\parallel}} {\sqrt{P^+_{\bot} P^-_{\bot}}} \
C^{\prime}_K(n_s) \ C^{\prime}_K(-n_s-4) \ k^{-4}_{\bot} \ ,
\label{pak4}
\end{eqnarray}

\vspace{0.4cm}
\noindent
where the last formulation shows the symmetry with respect to $s$.
The power laws of the spectra $E^s_{\parallel}$ have the same indices than 
those of $E^s_{\bot}$ and the Kolmogorov constant $C^{\prime}$ is in fact 
related to $C$ by the relation
\begin{equation}
{C^{\prime}_K(n_s) \over C^{\prime}_K(-n_s-4)} = {C_K(-n_s-4) \over C_K(n_s)} \, .
\end{equation}
Therefore the choice of the ratio $P^+_{\bot} / P^-_{\bot}$ determines not only
$C_K(n_s)$ but also $C^{\prime}_K(n_s)$, allowing for free choices of the 
dissipative rates of energy $P^{\pm}_{\parallel}$. 

\begin{figure}
\centerline{\psfig{file=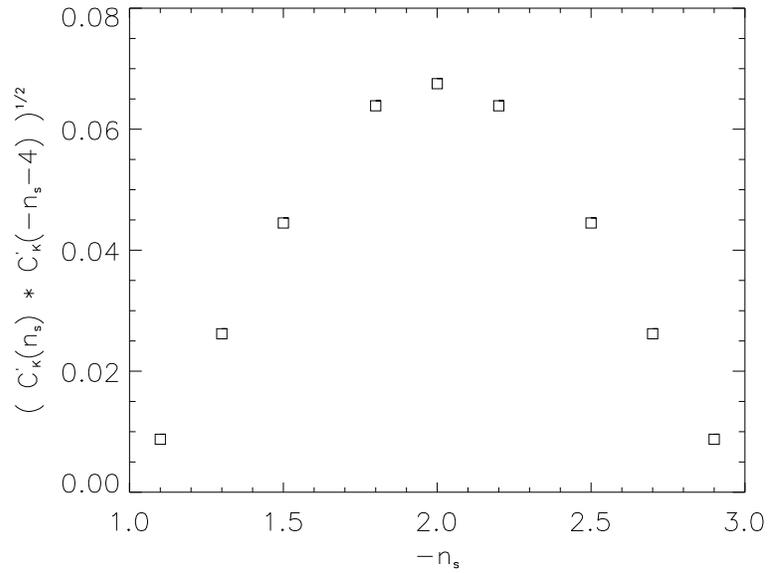,width=11cm,clip=}}
\caption[]{Variation of $\sqrt{C^{\prime}_K(n_s) \ C^{\prime}_K(-n_s-4)}$ as a 
function of $-n_s$. Notice the symmetry around the value $-n_s=2$ corresponding 
to the zero cross correlation case.}
\label{f3}
\end{figure}

The result of the numerical evaluation of $C^{\prime}_K(n_s)$ is shown in 
Figure \ref{f3}. 
An asymmetrical form is also visible\,; notice also that the values of 
$C^{\prime}_K(n_s)$ (\ie the constant in front of the parallel energy spectra)
are smaller by an order of magnitude than those of $C_K(n_s)$ for the
perpendicular spectra.

%%%%%%%%%%%%%%%%%%%%%%%%%%%%%%%%%%%%%%%%%%%%%%%%%%%%%%%%%%%%%%%%%%%%%%%%%%%%%%%%
\subsection{Temporal evolution of the kinetic equations}
\subsubsection{Numerical method}
%%%%%%%%%%%%%%%%%%%%%%%%%%%%%%%%%%%%%%%%%%%%%%%%%%%%%%%%%%%%%%%%%%%%%%%%%%%%%%%%

Equations (\ref{equ2D_1}) and (\ref{equ2D_2}) can be integrated numerically 
with a standard method, as for example presented in 
Leith and Kraichnan (1972).
%\cite{LK}. 
Since the
energy spectrum varies smoothly with $k$, it is convenient to use a logarithmic 
subdivision of the $k$ axis 
\begin{equation}
k_i = \delta k \ 2^{i/F} \ , 
\end{equation}
where $i$ is a non--negative integer\,; $\delta k$ is the minimum wave number in
the computation and $F$ is the number of wave numbers per octave. 
$F$ defines the refinement of the ``grid'', and in particular it is easily seen
that a given value of $F$ introduces a cut--off in the degree of non--locality
of the nonlinear interactions included in the numerical computation of the
kinetic equations. But since the solutions are local, a moderate value of $F$
can be used (namely, we take $F=4$). Tests have nevertheless been
performed with $F=8$ and we show that no significant changes occur in the 
results to be described below.

This technique allows us to reach Reynolds numbers much greater than in direct 
numerical simulations. In order to regularize the equations at large $k$, we 
have introduced dissipative terms which were omitted in the derivation of the 
kinetic equations. We take the magnetic Prandtl number ($\nu/\eta$) to be unity. 
For example, with $\delta k = 2^{-3}$, $F=8$, $i_{max}=225$\,; this corresponds 
to a ratio of scales $2^{28}/2^{-3}$. Taking a wave energy $U_0^2$ and an 
integral scale $L_0$ both of order one initially, and a kinematic viscosity of 
$\nu=3.3 \times 10^{-8}$, the Reynolds number of such a computation is 
${\cal R}_e=U_0\ L_0/\nu \sim 10^8$. 
All numerical simulations to be reported here have been computed on an Alpha 
Server 8200 located at the Observatoire de la C\^ote d'Azur (SIVAM).

%%%%%%%%%%%%%%%%%%%%%%%%%%%%%%%%%%%%%%%%%%%%%%%%%%%%%%%%%%%%%%%%%%%%%%%%%%%%%%%%
\subsubsection{Shear-Alfv\'en waves}
%%%%%%%%%%%%%%%%%%%%%%%%%%%%%%%%%%%%%%%%%%%%%%%%%%%%%%%%%%%%%%%%%%%%%%%%%%%%%%%%

\begin{figure}
\centerline{\psfig{file=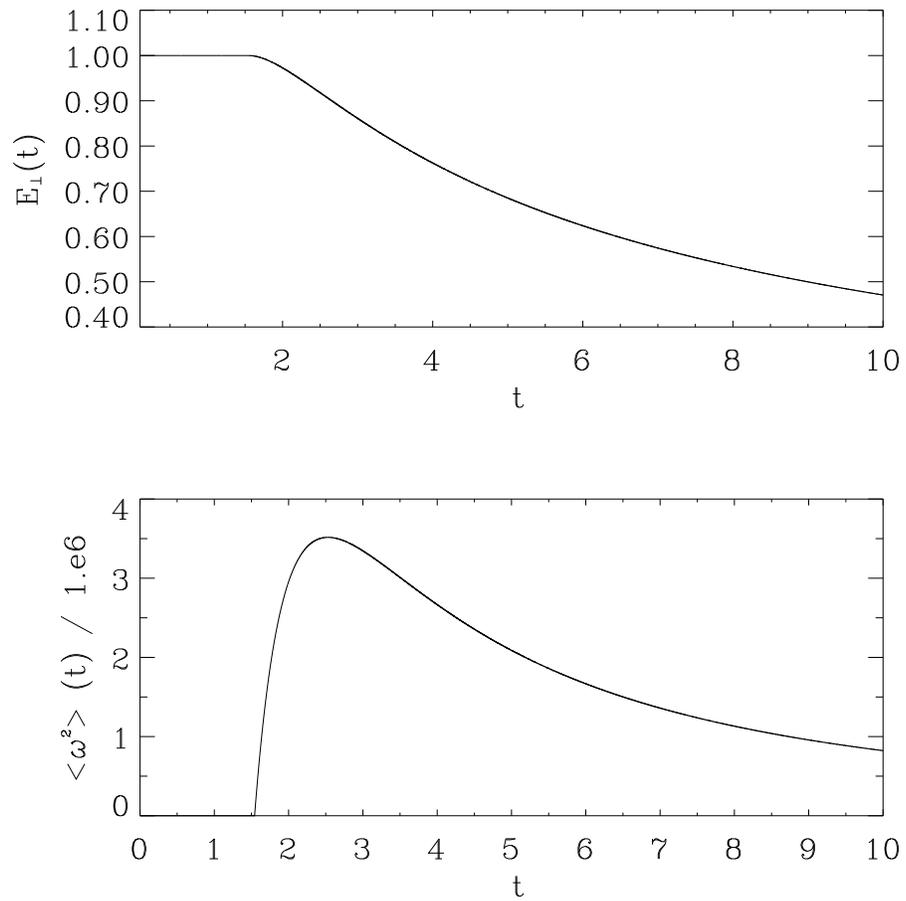,width=13cm,clip=}}
\caption[]{Temporal evolution of the energy $E_{\bot}(t)$ (top) and the enstrophy 
$\langle \omega^2(t) \rangle$ in units of $1. \, 10^6$. Notice the conservation 
of the energy up to the time $t_0 \simeq 1.55$.}
\label{f4}
\end{figure}

\begin{figure}
\centerline{\psfig{file=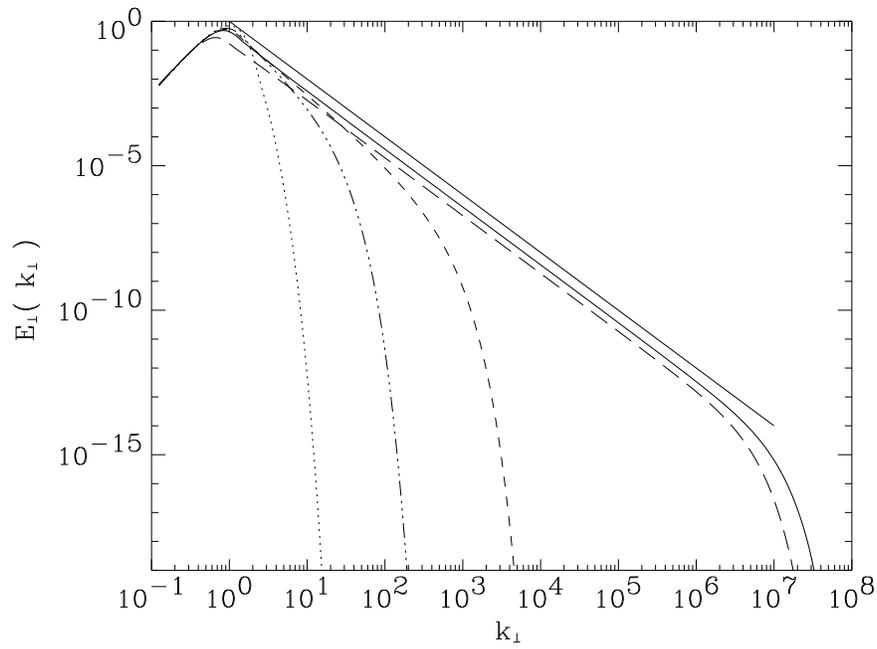,width=12cm,clip=}}
\caption[]{Energy spectra $E_{\bot}(k_{\bot},0)$ of the shear-Alfv\'en waves in 
the zero cross correlation case for the times $t=0$ (dot), $t=1.0$ (dash-dot), 
$t=1.5$ (small-dash), $t=1.6$ (solid) and $t=10.0$ (long-dash)\,; the straight 
line follows a $k^{-2}_{\bot}$.}
\label{f5}
\end{figure}

In this paper, we only consider decaying turbulence. 
As a first numerical simulation we have integrated the equation (\ref{equ2D_1}) 
in the zero cross--correlation case ($E^+=E^-$) and without forcing. 
Figure \ref{f4} (top) shows the temporal evolution of the total 
energy $E_{\bot}(t)$ with by definition 
\begin{equation}
E_{\bot}(t) = \int_{k_{min}}^{k_{max}} E_{\bot}(k_{\bot},0) \, d k_{\bot} \ ,
\end{equation}
where $k_{min}$ and $k_{max}$ have the values given in the previous section. 
%\begin{figure}
%\centerline{\psfig{file=Fig5_wt.ps,width=13cm,clip=}}
%\caption[]{Temporal evolution of the energy $E_{\bot}(t)$ (top) and the enstrophy 
%$\langle \omega^2(t) \rangle$ in units of $1. \, 10^6$. Notice the conservation 
%of the energy up to the time $t_0 \simeq 1.55$.}
%\label{f4}
%\end{figure}
The total energy is conserved up to a time $t_0 \simeq 1.55$ after  
which it decreases because of the dissipative effects linked to mode coupling, 
whereas the enstrophy $\int k^2 E_{\bot}(k) \, d^2{\bf k}$
increases sharply (bottom of Figure \ref{f4}). 
The energy spectra at different times are displayed in Figure \ref{f5}.
As we approach the time $t_0$, the spectra spread out to reach the smallest
scales (\ie the largest wavenumbers). For $t>t_0$, constant energy flux spectrum
$k^{-2}_{\bot}$ is obtained (indicated by the straight line). For times $t$ 
significantly greater than $t_0$, we have a self-similar energy decay, in what 
constitutes the turbulent regime. 
%\begin{figure}
%\centerline{\psfig{file=Fig6_wt.ps,width=12cm,clip=}}
%\caption[]{Energy spectra $E_{\bot}(k_{\bot},0)$ of the shear-Alfv\'en waves in 
%the zero cross correlation case for the times $t=0$ (dot), $t=1.0$ (dash-dot), 
%$t=1.5$ (small-dash), $t=1.6$ (solid) and $t=10.0$ (long-dash)\,; the straight 
%line follows a $k^{-2}_{\bot}$.}
%\label{f5}
%\end{figure}

%%%%%%%%%%%%%%%%%%%%%%%%%%%%%%%%%%%%%%%%%%%%%%%%%%%%%%%%%%%%%%%%%%%%%%%%%%%%%%%%
\subsubsection{Shear-Alfv\'en {\it versus} pseudo-Alfv\'en waves}
%%%%%%%%%%%%%%%%%%%%%%%%%%%%%%%%%%%%%%%%%%%%%%%%%%%%%%%%%%%%%%%%%%%%%%%%%%%%%%%%

\begin{figure}
\centerline{\psfig{file=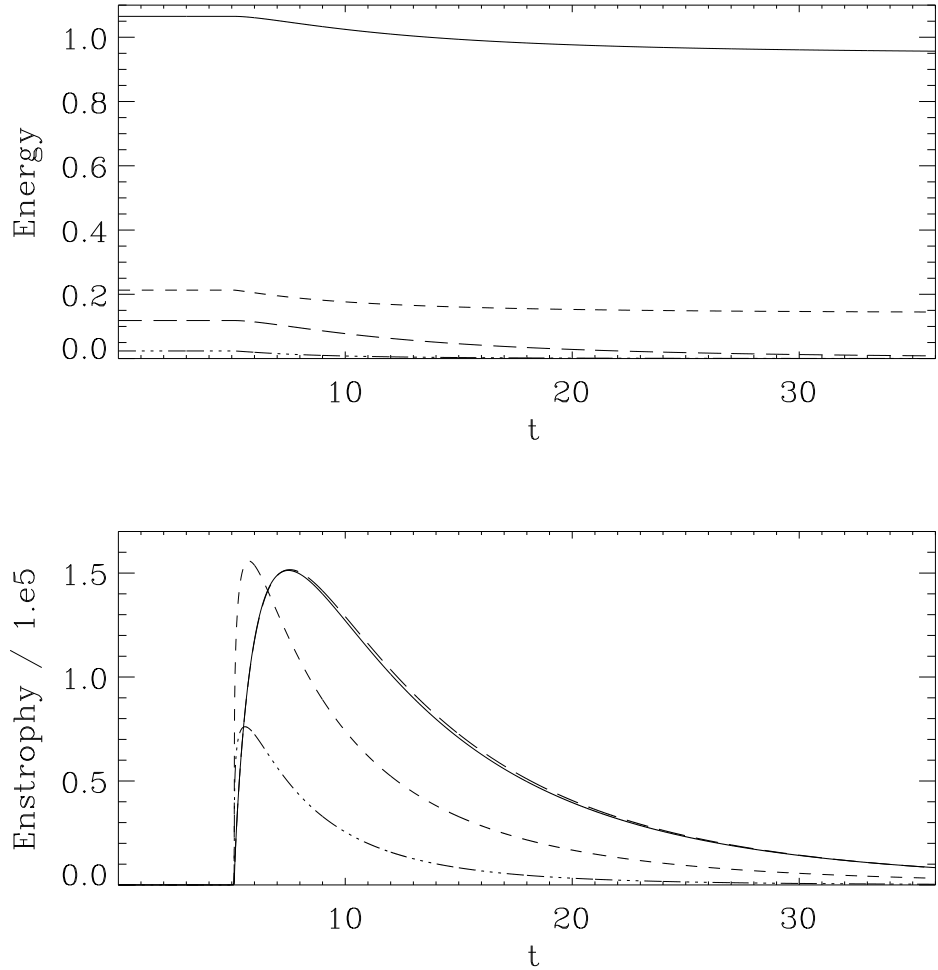,width=14cm,clip=}}
\caption[]{Temporal evolution of energies (top) $E^+_{\bot}$ (solid), $E^-_{\bot}$ 
(long-dash), $E^+_{\parallel}$ (small-dash) and $E^-_{\parallel}$ (dash-dot)\,;
the same notation is used for the enstrophies (bottom) which are in units of 
$1. \, 10^5$. Notice that energies are conserved till the time 
$t_0^{\prime} \simeq 5$.}
\label{f6}
\end{figure}

\begin{figure}
\centerline{\psfig{file=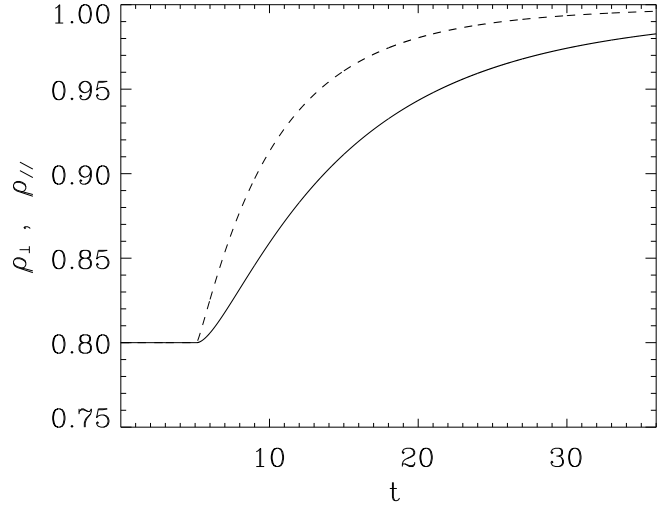,width=10cm,clip=}}
\caption[]{Temporal evolution of the cross correlations $\rho_{\bot}$ (solid) and 
$\rho_{\parallel}$ (dash). These quantities are conserved up to the time 
$t_0^{\prime}$.}
\label{f7}
\end{figure}

\begin{figure}
\centerline{\psfig{file=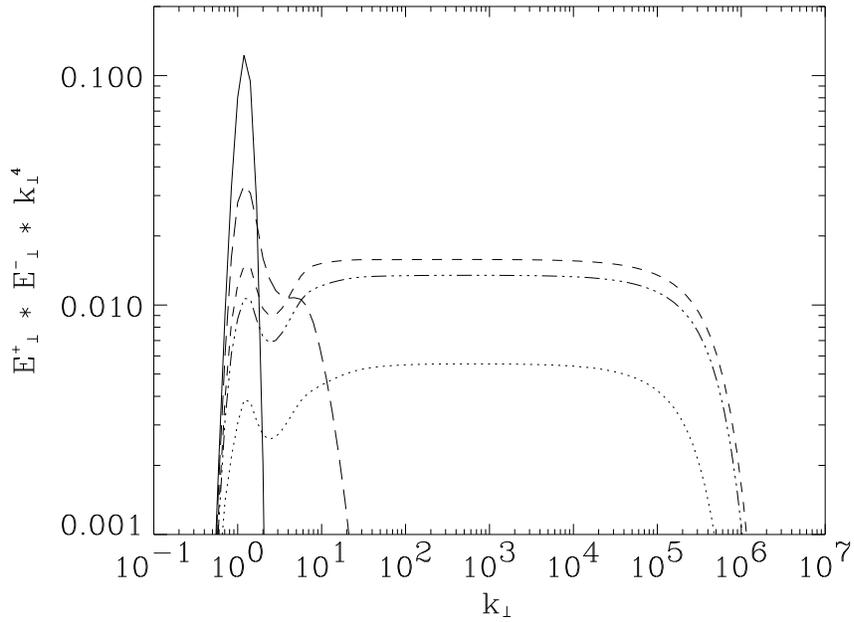,width=12cm,clip=}}
\caption[]{Compensated spectra $E^+_{\bot} E^-_{\bot} k^4_{\bot}$ at times $t=0$ 
(solid), $t=4$ (long-dash), $t=6$ (small-dash), $t=8$ (dash-dot) and $t=20$ (dot).}
\label{f8}
\end{figure}

\begin{figure}
\centerline{\psfig{file=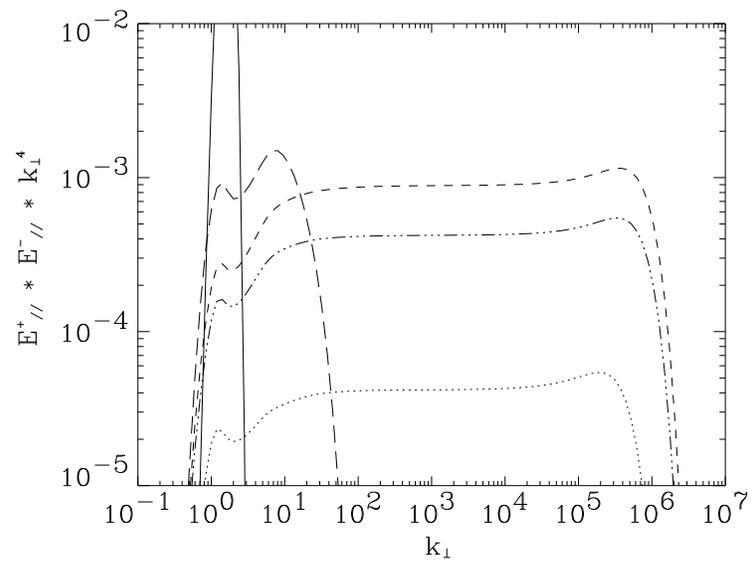,width=10.5cm,clip=}}
\caption[]{Compensated spectra $E^+_{\parallel} E^-_{\parallel} k^4_{\bot}$ for 
the same times and with the same symbols as in Figure \ref{f8}.}
\label{f9}
\end{figure}

%\begin{figure}
%\centerline{\psfig{file=Fig7_wt.ps,width=14cm,clip=}}
%\caption[]{Temporal evolution of energies (top) $E^+_{\bot}$ (solid), $E^-_{\bot}$ 
%(long-dash), $E^+_{\parallel}$ (small-dash) and $E^-_{\parallel}$ (dash-dot)\,;
%the same notation is used for the enstrophies (bottom) which are in units of 
%$1. \, 10^5$. Notice that energies are conserved till the time 
%$t_0^{\prime} \simeq 5$.}
%\label{f6}
%\end{figure}
%\begin{figure}
%\centerline{\psfig{file=Fig8_wt.ps,width=10cm,clip=}}
%\caption[]{Temporal evolution of the cross correlations $\rho_{\bot}$ (solid) and 
%$\rho_{\parallel}$ (dash). These quantities are conserved up to the time 
%$t_0^{\prime}$.}
%\label{f7}
%\end{figure}
In a second numerical computation we have studied the system (\ref{equ2D_1}) 
(\ref{equ2D_2}) with an initial normalised cross--correlation of $80 \%$. 
The following parameters have been used\,: $\delta k = 2^{-3}$, $F=4$, 
$i_{max}=105$ and $\nu=6.4 \,10^{-8}$. Figure \ref{f6} (top)
shows the temporal evolution of energies for the four different waves 
($E^{\pm}_{\bot}$ and $E^{\pm}_{\parallel}$). The same behavior as that of 
Figure \ref{f4} (top) is observed, with a conservation of energy up to the time 
$t_0^{\prime} \simeq 5$, and a decay afterwards\,; this decay is nevertheless
substantially weaker than when the correlation is zero, since in the presence of
a significant amount of correlation between the velocity and the magnetic field,
it is easily seen from the primitive MHD equations that the nonlinearities are
strongly reduced. On the other hand the temporal evolution of 
enstrophies (bottom) displays that the maxima for these four types of waves
are reached at different times\,: the pseudo-Alfv\'en waves are the fastest 
to reach their maxima at $t \simeq 5.5$ 
\vs $t \simeq 7.5$ for the shear-Alfv\'en waves.
Figure \ref{f7} corresponds to the temporal evolution of another conserved 
quantity, the cross correlation $\rho_x$ defined as 
\begin{equation}
\rho_x = {E^+_x - E^-_x \over E^+_x + E^-_x} \ ,
\end{equation}
where $x$ symbolizes either $\bot$ or $\parallel$. As expected, $\rho_x$ is 
constant during an initial period (till $t=t_0^{\prime}$) and then tends 
asymptotically to one, but in a faster way for the pseudo-Alfv\'en waves. 
This growth of correlation is well documented in the isotropic case
(Matthaeus and Montgomery 1980)
and is seen to hold as well here in the weak turbulence regime.
Figures \ref{f8} and \ref{f9} give the compensated spectra 
$E^+_{\bot} E^-_{\bot} k^4_{\bot}$
and $E^+_{\parallel} E^-_{\parallel} k^4_{\bot}$ respectively  at different 
times. In both cases, from $t=6$ onward, a plateau is observed over almost four 
decades and remains flat for long times\,; this illustrates nicely the 
theoretical predictions (\ref{pek4}) and (\ref{pak4}). 

%\begin{figure}
%\centerline{\psfig{file=Fig9_wt.ps,width=12cm,clip=}}
%\caption[]{Compensated spectra $E^+_{\bot} E^-_{\bot} k^4_{\bot}$ at times $t=0$ 
%(solid), $t=4$ (long-dash), $t=6$ (small-dash), $t=8$ (dash-dot) and $t=20$ (dot).}
%\label{f8}
%\end{figure}
%\begin{figure}
%\centerline{\psfig{file=Fig10_wt.ps,width=10.5cm,clip=}}
%\caption[]{Compensated spectra $E^+_{\parallel} E^-_{\parallel} k^4_{\bot}$ for 
%the same times and with the same symbols as in Figure \ref{f8}.}
%\label{f9}
%\end{figure}

%%%%%%%%%%%%%%%%%%%%%%%%%%%%%%%%%%%%%%%%%%%%%%%%%%%%%%%%%%%%%%%%%%%%%%%%%%%%%%%%
\section{Front propagation}
%%%%%%%%%%%%%%%%%%%%%%%%%%%%%%%%%%%%%%%%%%%%%%%%%%%%%%%%%%%%%%%%%%%%%%%%%%%%%%%%

\begin{figure}
\centerline{\psfig{file=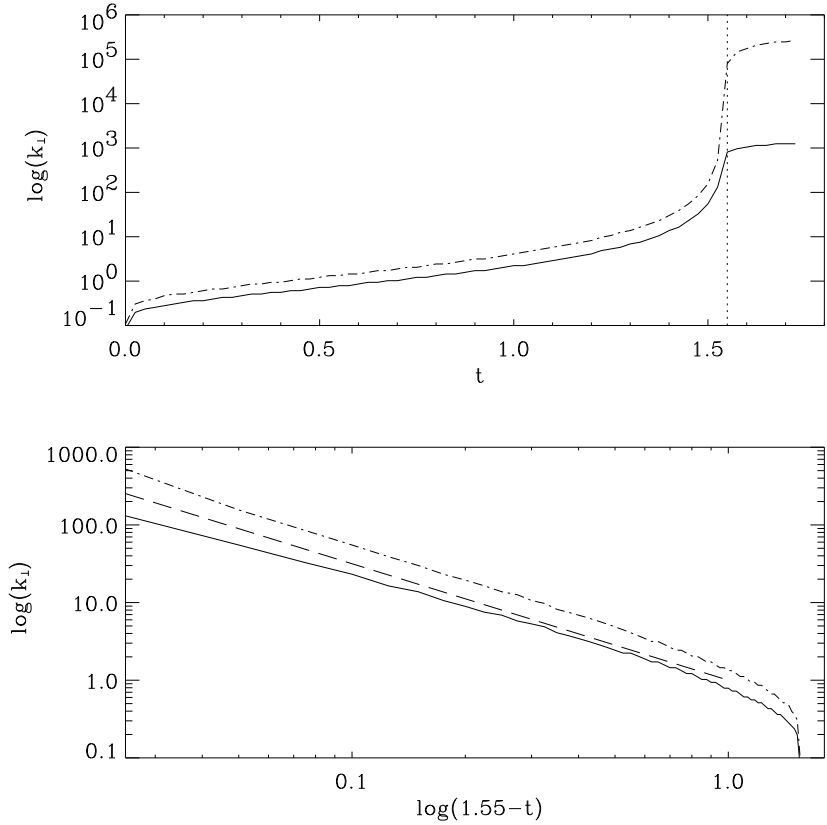,width=11.5cm,clip=}}
\caption[]{Temporal evolution (top), in lin-log coordinates, of the front of 
energy propagating to small scales. The solid line and the dash-dot line 
correspond respectively to an energy of $10^{-16}$ and $10^{-25}$. An abrupt 
change is visible at time $t_0 \simeq 1.55$ (vertical dotted line). The 
$\log(\kpe)$ as a function of $\log(1.55-t)$ (bottom) displays a power law in 
$\kpe \sim (1.55-t)^{-1.5}$ (large dash line).}
\label{front}
\end{figure}

\begin{figure}
\centerline{\psfig{file=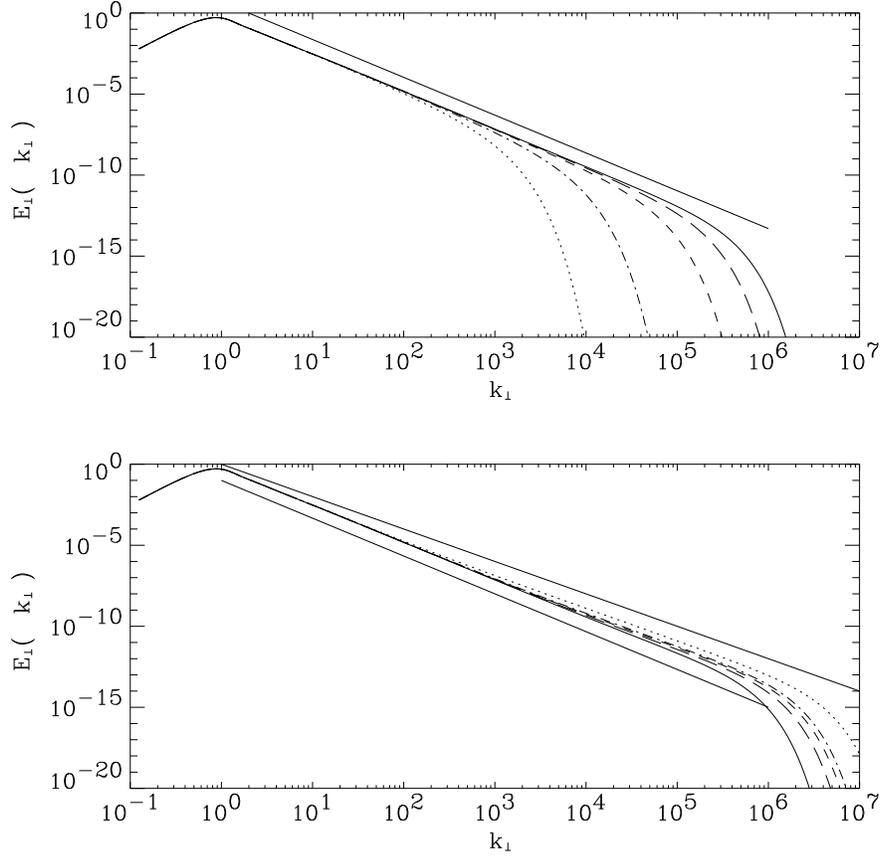,width=12.cm,clip=}}
\caption[]{Temporal evolution of the energy spectrum $E_{\bot}(k_{\bot},0)$ of 
the shear-Alfv\'en waves around the catastrophic time $t_0 \simeq 1.544$. 
For $t < t_0$ (top), $t=1.50$ (dot), $t=1.53$ (dash-dot), $t=1.54$ (dash),
$t=1.542$ (long dash) and $t=1.543$ (solid)) a $k_{\perp}^{-7/3}$-- spectrum is 
observed.
For $t \ge t_0$ (bottom), $t=1.544$ (solid), $t=1.546$ (long dash), $t=1.548$ 
(dash), $t=1.55$ (dash-dot) and $t=1.58$ (dot)) a fast change of the slope 
appears to give finally a $k_{\perp}^{-2}$-- spectrum. Note that this change
propagates from small scales to large scales. In both cases straight lines 
follow either a $k^{-7/3}_{\bot}$ or a $k^{-2}_{\bot}$.}
\label{k73}
\end{figure}

The numerical study of the transition between the initial state and the final 
state, where the $k_{\perp}^{-2}$--spectrum is reached, shows two remarkable 
properties illustrated by Figure \ref{front} and \ref{k73}. 

We show in Figure \ref{front} (top), in lin-log coordinates, the progression with 
time of the front of energy propagating to small scales\,; more precisely, we 
give the wavenumber at time $t$ with an energy of, respectively, $10^{-25}$ 
(dash-dot line) and $10^{-16}$ (solid line). 
Note that all curves display an abrupt 
change at $t_0 \simeq 1.55$, after which the growth is considerably slowed down. 
Using this data, Figure \ref{front} (bottom) gives $\log(\kpe)$ as a function of 
$\log(1.55-t)$, the lines having the same meaning as in Figure \ref{front} (top);
the large dash represents a power law $\kpe \sim (1.55-t)^{-1.5}$. Hence, 
the small scales, in this weak turbulence formalism, are reached in a finite time
\ie in a catastrophic way. This is also seen on the temporal evolution of the 
enstrophy (see bottom of Figure \ref{f4}), with a catastrophic growth ending at
$t \simeq 2.5$, after which the decay of energy begins.
%\begin{figure}
%\centerline{\psfig{file=Fig11_wt.ps,width=11.5cm,clip=}}
%\caption[]{Temporal evolution (top), in lin-log coordinates, of the front of 
%energy propagating to small scales. The solid line and the dash-dot line 
%correspond respectively to an energy of $10^{-16}$ and $10^{-25}$. An abrupt 
%change is visible at time $t_0 \simeq 1.55$ (vertical dotted line). The 
%$\log(\kpe)$ as a function of $\log(1.55-t)$ (bottom) displays a power law in 
%$\kpe \sim (1.55-t)^{-1.5}$ (large dash line).}
%\label{front}
%\end{figure}
%\begin{figure}
%\centerline{\psfig{file=Fig12_wt.ps,width=12.cm,clip=}}
%\caption[]{Temporal evolution of the energy spectrum $E_{\bot}(k_{\bot},0)$ of 
%the shear-Alfv\'en waves around the catastrophic time $t_0 \simeq 1.544$. 
%For $t < t_0$ (top), $t=1.50$ (dot), $t=1.53$ (dash-dot), $t=1.54$ (dash),
%$t=1.542$ (long dash) and $t=1.543$ (solid)) a $k_{\perp}^{-7/3}$-- spectrum is 
%observed.
%For $t \ge t_0$ (bottom), $t=1.544$ (solid), $t=1.546$ (long dash), $t=1.548$ 
%(dash), $t=1.55$ (dash-dot) and $t=1.58$ (dot)) a fast change of the slope 
%appears to give finally a $k_{\perp}^{-2}$-- spectrum. Note that this change
%propagates from small scales to large scales. In both cases straight lines 
%follow either a $k^{-7/3}_{\bot}$ or a $k^{-2}_{\bot}$.}
%\label{k73}
%\end{figure}

Figure \ref{k73} shows the temporal evolution of the energy spectrum 
$E_{\bot}(k_{\bot},0)$ of the shear-Alfv\'en waves around the catastrophic time
$t_0$. We see that before $t_0$, evaluated here with a better precision to be
equal to
$1.544$, the energy spectrum propagates to small scales following a 
stationary $k_{\perp}^{-7/3}$-- spectrum and not a $k_{\perp}^{-2}$-- spectrum.
It is only when the dissipative scale is reached, at $t_0$, that a remarkable 
effect is observed\,: in a very fast time the $k_{\perp}^{-7/3}$ solution turns 
into the finite energy flux spectrum $k_{\perp}^{-2}$ with a change of the slope 
propagating from small scales to large scales. 

Note that this picture is different from the scenario proposed by Falkovich and 
Shafarenko (1991, hereafter FS) for the finite capacity spectra. In an example
considered by FS, 
the Kolmogorov spectrum forms right behing the propagating front, whereas in
our case it forms only after the front reaches infinite wavenumbers (\ie 
dissipative region). The front propagation can be described in terms of 
self-similar solutions having a form (Falkovich and Shafarenko 1991; Zakharov et 
al. 1992): 
\begin{equation}
E_{\bot}(k_{\bot},0) = {1 \over \tau^a} E_0 ({k_{\perp} \over \tau^b}) \, ,
\label{efront}
\end{equation}
where $\tau = t_0 - t$. Substituting (\ref{efront}) into the kinetic equation
(\ref{equ2D_1}) we have 
$$
\partial_{\tau} \left({1 \over \tau^a} E_0 ({k_{\perp} \over \tau^b}) \right) 
\sim \tau^{-a} E_0 ({\kappa_{\perp} \over \tau^b}) \left( \tau^{b-a} 
E_0 ({L_{\perp} \over \tau^b}) - \tau^{b-a} E_0 ({k_{\perp} \over \tau^b}) 
\right) \tau^{2b} \, .
$$
which leads to the relation
\begin{equation}
1 + 3b = a \, .
\label{efront2}
\end{equation}
If $E_0$ is stationary and has a power-law form $E_0 \sim k^{m}$, then we have 
another relation between $a$ and $b$
\begin{equation}
a + mb = 0
\label{efront4}
\end{equation}
Excluding $a$ from (\ref{efront2}) and (\ref{efront4}) we have $1+(3+m)b=0$.
In our case this condition is satisfied because $b = -3/2$ and $m=-7/3$ which 
confirms that the front solution is of self-similar type.

%%%%%%%%%%%%%%%%%%%%%%%%%%%%%%%%%%%%%%%%%%%%%%%%%%%%%%%%%%%%%%%%%%%%%%%%%%%%%%%%
\section{Locality  of  power-law spectra}
\label{var}
%%%%%%%%%%%%%%%%%%%%%%%%%%%%%%%%%%%%%%%%%%%%%%%%%%%%%%%%%%%%%%%%%%%%%%%%%%%%%%%%

As we mentioned above, a Kolmogorov-type spectrum obtained {\it via} the
Zakharov transform is a solution to the kinetic equations if,
and only if, the original collision integral in this equation
(before the Zakharov transformation) converges on it, -- a property
called locality of the spectrum.
Having in mind that the front propagation spectrum is also of
a power law type, let us study locality of power law spectra of a general form, 
\begin{equation}
 E^s_{\bot} (k_{\bot},0) = k_{\bot}^{m_s} , \;\;\;\; 
 E^{-s}_{\bot} (k_{\bot},0) = k_{\bot}^{m_{-s}},
\label{pls}
\end{equation}
where indices $m_s$ and $m_{-s}$ are arbitrary numbers.
Recall that the collision integral in (\ref{equ2D_1}) is to be taken over a 
semi-infinite strip shown in Figure \ref{fzakh}. It may be singular only at the 
following three  points, 

(p1) \, $\kappa_\perp = L_\perp = + \infty, $

(p2) \, $\kappa_\perp = k_\perp, \;\;\; L_\perp = 0, $

(p3) \, $\kappa_\perp = 0, \;\;\; L_\perp = k_\perp,$

i.e. the corners and infinity of the integration area shown in
Figure 1. To study convergence at point (p1) it is convenient to
change variables,
\begin{equation}
\kappa_\perp + L_\perp=r_+, \;\;\;\; 
\kappa_\perp - L_\perp=r_-, \;\;\;\;  k_\perp <r_+ < + \infty,
\;\;\;\;  -k_\perp <r_- <  k_\perp.
\end{equation}
Taking the limit $r_+ \to + \infty$ in the integrand (which
corresponds to (p1)) and retaining the largest terms, we obtain
the following conditions of convergence,
\begin{equation}
m_{-s} + m_{s} <0, \;\;\;\; m_{-s} <-1.
\end{equation}
In the vicinity of (p2) it is convenient to use the polar 
coordinates,
\begin{equation}
\kappa_\perp = k_\perp (1+ r \, \cos \theta), \;\;\;\; L_\perp = k_\perp r \, 
\sin \theta, \;\;\;\; - {\pi \over 4} < \theta < {\pi \over 4}, \;\;\;\;  
-k_\perp <r <  k_\perp.
\end{equation}
Considering the limit $r \to 0$ and integrating over $\theta$ one can see that the 
collision integral converges if, and only if, $ m_{s} >-3$. Similarly, one obtains 
the convergence condition at point (p3) which is $ m_{-s} >-3$.

All the convergence conditions in the kinetic equation
for $E^{-s}$ are, of course, symmetric to the case of 
$E^{s}$; one simply has to exchange $ m_{-s}$ and $ m_{s}$
in these conditions. Summarizing, one can write the
conditions for simultaneous convergence for both
$E^{s}$ and $E^{-s}$,
\begin{equation}
-3 < m_{\pm} < -1. 
\end{equation}
The Kolmogorov spectral exponents lie on the line $m_++ m_-= -4$,
and the locality interval $(-3,-1)$ for one of them maps exactly
onto the same interval for another exponent.
In particular, the symmetric $-2$ Kolmogorov spectum is local.
One can also see that the front solution with index $-7/3$ is local
according to the above locality condition.

%%%%%%%%%%%%%%%%%%%%%%%%%%%%%%%%%%%%%%%%%%%%%%%%%%%%%%%%%%%%%%%%%%%%%%%%%%%%%%%%
\section{Fokker-Planck approximation}
%%%%%%%%%%%%%%%%%%%%%%%%%%%%%%%%%%%%%%%%%%%%%%%%%%%%%%%%%%%%%%%%%%%%%%%%%%%%%%%%

In the previous section, we established the fact that both
the Kolmogorov $-2$ and front $-7/3$ spectra are local.
However, during the initial phase of the turbulence
decay, turbulence may be very nonlocal. Namely,  the
nonlinear  interaction for short waves will be dominated by triads
that involve a long wave corresponding to the initial
large-scale turbulence. Our locality analysis suggests that
this will happen when the slope of the large-scale part of the spectrum
is still steeper than $-3$, i.e. neither a $-7/3$ nor a $-2$ small-scale
tail has grown strong enough in amplitude yet for the local interaction to
take over. Further, the locality analysis suggests that
the dominant contribution to the collision integral in this
case will come from small vicinities of the points (p2) and (p3)
both of which involve one small wavenumber\,: $L$ and $\kappa$
correspondingly. Thus, one can expand the integrand of the
collision integral in powers of these wavenumbers and
reduce the kinetic equation to a  second order Fokker-Planck equation,
similarly to the way it was done for the Rossby three-wave process
in Balk et al. (1990a, 1990b). 
Below, we will derive such an equation considering contributions
of the points (p2) and (p3) separately.

Below in this section, we will consider only the two-dimensional symmetric case 
and, therefore, we omit the superscript $s$ in $E^s$ and the subscript 
$\perp$ for the 
wavevectors. The kinetic equation (\ref{equ2D_1}) can be rewritten as
\begin{equation}
\frac{\partial E (k)}{\partial t} = 2 \int
F(k, \kappa, L) d \kappa dL,
\label{kinF}
\end{equation}
where
\begin{equation}
F(k, \kappa, L) =
{\pi \varepsilon^2 \over 2 b_0}  (\cos \phi)^2 \, |sin \phi| \ 
\frac{k^2L^2}{\kappa^2}
\ E (\kappa) \ \Bigg[  
{E (L) \over L}  - {E (k) \over k}\Bigg],
 \label{Fkl}
\end{equation}
and $\phi$ is an angle between wavevectors $\bf k$ and $L$, so that
\begin{equation}
\cos \phi ={k^2 +L^2 -\kappa^2 \over 2 k L}.
\label{cosphi}
\end{equation}

In a small vicinity of (p2) one can expand 
$F$  in powers of small
$L$ and $h=k - \kappa=O(L)$.
Taking into account that
$\cos^2 \phi = (h/L)^2 +O(L)$, we have
\begin{equation}
F(k, \kappa, L) =
{\pi \varepsilon^2 \over 2 b_0}  (h/L)^2 \, |1-(h/L)^2|^{1/2} L
\, E (k) E (L).
 \label{Fp2}
\end{equation}
Substituting this expression into (\ref{kinF}) and integrating over
$h$ from $-L$ to $L$ we have the following contribution of the point (p2),
\begin{equation}
\Big( \frac{\partial E (k)}{\partial t} \Big)_2 = 2D \, E(k),
 \label{cont2}
\end{equation}
where the constant $D$ is
\begin{equation}
D= {\pi^2 \epsilon^2 \over 16 b_0} \int_0^\infty L^2 E(L) \, dL.
\label{Dc}
\end{equation}

Let us consider now the contribution to the 
collision integral that comes from the vicinity of the point (p3).
Introducing the new variable $l$ so that $L=k+l$ and applying the Zakharov
transformation (simply $l \to -l$ near point (p3)), we
rewrite (\ref{kinF}) as follows
\begin{equation}
\frac{\partial E (k)}{\partial t} = \int_0^\infty d \kappa
\int_{-\kappa}^{\kappa} dl [F(k, \kappa, k+l) + F(k, \kappa, k-l)].
\label{+-l}
\end{equation}
Assuming that $\kappa$ and $l$ are small and  that they are of 
the same order near point (p3) we have
\begin{eqnarray}
{E (k+l) \over k+l}  - {E (k) \over k} & =& l \, \Bigg[   
\frac{\partial (E (k')/k')}{\partial k'} \Bigg]_{k'=k+l/2} + O(l^3), \\
k^2L^2 = k^2 (k+l)^2  & =& (k+l/2)^4 +O(l^2), \\
\cos \phi & =& 1 + O(l^2), \\
|\sin \phi| & =&  \sqrt{|l^2 - \kappa^2|} ((k+l/2)^{-1} + O(l^2)).
\end{eqnarray}
Substituting these expressions into (\ref{+-l}) and further
Taylor expanding the integrand and integrating over $l$
we have the following main order contribution of the point (p3),
\begin{equation}
\Big(\frac{\partial E (k)}{\partial t} \Big)_3 
= D \, {\partial \over \partial k}
\big( k^3 {\partial \over \partial k} \big({ E (k) \over k} \big) \big),
\label{cont3}
\end{equation}
where the diffusion constant $D$ is given by (\ref{Dc}).
Combining (\ref{cont2}) and  (\ref{cont3}) we have the following
kinetic Fokker-Planck equation,
\begin{equation}
\frac{\partial E (k)}{\partial t} 
= D \, {\partial \over \partial k}
\big( k^3 {\partial \over \partial k} \big({ E (k) \over k} \big) \big)
+2D  E (k).
\label{f-planck}
\end{equation}
The first term in the RHS of this equation conserves energy
of the short waves because the three-wave interaction
near (p3) does not transfer any energy to (or from) the 
large-scale component. The second term is not conservative\,:
it describes a direct nonlocal energy transfer from long
waves to the short ones. According to (\ref{f-planck}),
the total energy of short waves grows exponentially. Indeed
one can rewrite this equation in the form of a local
conservation law for $N=e^{-2Dt} E$ as follows,
\begin{equation}
\frac{\partial N}{\partial t} 
= D \, {\partial \over \partial k}
\big( k^3 {\partial \over \partial k} \big({ N \over k} \big) \big).
\label{f-pN}
\end{equation}
It is interesting that this equation can also be rewritten
in the form of a conservation for $N/k^2$,
\begin{equation}
\frac{\partial (N/k^2)}{\partial t} 
= D \, {\partial \over \partial k}
\big( k^{-1} {\partial \over \partial k} ({k N }) \big).
\label{f-pN2}
\end{equation}
Further, there are two independent power law solutions to
(\ref{f-pN}) and (\ref{f-pN2})\,: $N \sim k$ and $N \sim 1/k$.
The first of these solutions corresponds to the 
equipartition of $N$ and a constant flux of $N/k^2$, whereas
the second one corresponds to 
the equipartition of $N/k^2$ and a constant flux of $N$.
This property of the Fokker-Planck equations to have two
independent integrals of motion such that the constant flux
of one of them corresponds to the equipartition of another
one (and vice versa) was recently noticed in 
Nazarenko and Laval (1998)
%\cite{nl} 
in the context of the problem of passive scalars.
Note however, that one could expect  solutions $N \sim k^{\pm 1}$
only in a very idealized situation when a short wave turbulence
is generated by a source separated from the intense long
waves by a spectral gap, and only for a limited time until
the $-7/3$ tail growing from the large-scale side
will fill this spectral gap. In general, the dynamics given by
the Fokker-Planck equation (\ref{f-planck}) describes more
complex combination of the instability and diffusion processes
with an energy in-flux from the initial large scales.

%%%%%%%%%%%%%%%%%%%%%%%%%%%%%%%%%%%%%%%%%%%%%%%%%%%%%%%%%%%%%%%%%%%%%%%%%%%%%%%%
\section{MHD turbulence without an external magnetic field}
\label{swt5}
%%%%%%%%%%%%%%%%%%%%%%%%%%%%%%%%%%%%%%%%%%%%%%%%%%%%%%%%%%%%%%%%%%%%%%%%%%%%%%%%

We have considered until now a turbulence of Alfv\'en waves that arises in the
presence of a strong uniform magnetic field. Following 
Kraichnan (1965), 
%\cite{kr65},
one can assume that the results obtained for turbulence in
a strong external magnetic field are applicable to
MHD turbulence at small scales which experience the
magnetic field of the large-scale component as a 
quasi-uniform external field. Furthermore, the large-scale magnetic field is
much stronger than the one produced by the small-scales themselves
because most of the MHD energy is condensed at large scales due to the
decreasing distribution of energy among modes as the wavenumbers grow. In this 
case therefore, the small-scale dynamics consists again of a large number
of weakly interacting Alfv\'en waves.
Using such a  hypothesis and applying a dimensional argument, 
Kraichnan derived the $k^{-3/2}$ energy spectrum for MHD turbulence. However, 
Kraichnan did not take into account the local anisotropy associated with the 
presence of this external field. In Ng and Bhattacharjee (1997) (see also 
Goldreich and Sridhar 1997) the dimensional argument of Kraichnan is modified 
in order to take into account the anisotropic dependence of the characteristic 
time associated with Alfv\'en waves on the wavevector by simply writing
\begin{equation}
\tau \sim {1 \over b_0 k_{\pa}} \ .
\end{equation}
In that way, one obtains a
$k_\perp^{-2}$ energy spectrum, which agrees with the analytical and
numerical results of the present paper for the spectral 
dependence on $k_\perp$. On the other hand, the dependence of the spectra on 
$k_{\pa}$, as we showed before in this paper, is not  universal
because of the absence of energy transfer in the $k_{\pa}$ direction, although 
it is shown in 
Kinney and McWilliams (1998) 
%\cite{kima} 
that for a quasi--uniform field as 
considered in this section, there is some transfer in the quasi--parallel 
direction.
In the strictly uniform case, this spectral dependence is determined only
by the dependence on $k_{\pa}$ of the driving and/or initial conditions. 

For large time, the spectrum is almost two-dimensional.
The characteristic width of the spectrum in $k_{\pa}$ (described by the
function $f_1(k_\pa)$) is much less than its 
width in $k_\perp$, so that approximately one can write
\begin{equation}
e^s({\bf k})= C \, k_\perp^{-3} \, \delta(k_{\pa}),
\end{equation}
where $C$ is a constant.
The $k_\perp^{-3}$ factor corresponds to the 
$E_\perp \propto  k_\perp^{-2}$ Kolmogorov--like spectrum found in this paper
(the physical dimensions of $e^s$ and $E^s_\perp$ being different).
In the context of MHD turbulence, this spectrum is valid only locally,
that is for distances smaller than the length--scale of the
magnetic field associated with the energy--containing part of
MHD turbulence. Let us average this spectrum over the large 
energy containing scales, that is over all possible directions of ${\bf B_0}$.
Writing $k_\perp = |{\bf k} \times {\bf B_0}|/|B_0|$ and 
$k_{\pa} = |{\bf k} \cdot B_0|/|B_0|$ and assuming that ${\bf B_0}$ takes
all possible directions in 3D space with equal probability,
we have for the averaged spectrum
\begin{equation}
\langle e^s \rangle = \int e^s({\bf k}, {\bf x}) \, d\sigma(\zeta) =  
\int C \delta(\zeta \cdot {\bf k}) 
\, |\zeta \times {\bf k}|^{-3}  d\sigma(\zeta),
\end{equation}
where $\zeta =(\sin \theta \, \cos \phi,\sin \theta \, \sin \phi, \cos \theta)$
is a unit vector in the coordinate space and  
$d\sigma = \sin \theta \, d \theta \,d \phi$ is the
surface element on the  unit sphere.
Choosing  $\theta$ to be the angle between ${\bf k}$ and ${\bf B_0}$
and $\phi$ to be the angle
in the transverse to ${\bf k}$ plane, we have
\begin{equation}
\langle e^s \rangle = C \int_0^{2\pi} d \phi \int_0^\pi {\delta (\cos \theta) 
\over |k|} {(\sin \theta)^{-3} \over |k|^3} \, \sin \theta \, d \theta = 2 \pi 
C k^{-4}.
\end{equation}
This isotropic spectrum represents the averaged energy density in 3D wavevector 
space. By averaging over all possible directions of the wavevector,
we get the following density of the energy distribution over the absolute 
value of the wavevector,
\begin{equation}
E_k= 8 \pi^2 C \, k^{-2}.
\end{equation}
As we see, taking into account the local anisotropy and subsequent
averaging over the isotropic energy containing scales results in
an isotropic energy spectrum $k^{-2}$. This result is different 
from the $k^{-3/2}$ spectrum derived by Kraichnan without taking
into account the local anisotropy of small scales.
The difference in spectral indices may also arise from the fact that the 
approach here is that of weak turbulence, whereas in the strong turbulence case, 
isotropy is recovered on average and a different spectrum -- that of Kraichnan 
-- obtains. 

Solar wind data (Matthaeus and Goldstein 1982) indicates that the isotropic 
spectrum scales as $k^{-\alpha}$ with $\alpha\sim 1.67$, close to the Kolmogorov 
value for neutral fluids (without intermittency corrections which are known to 
occur)\,; hence, it could be interpreted as well as being a Kraichnan--like 
spectrum steepened by intermittency effects which are known to take place in 
strong MHD turbulence as well in the form of current and vorticity filaments,
ribbons and sheets, and magnetic flux tubes. 
However in the context of the interstellar medium (ISM), data show that the 
velocity dispersion is correlated with the size of the region observed (Larson 
1981; Scalo 1984; Falgarone et al. 1992). These correlations are approximately 
of power-law form such that the corresponding energy spectrum scales with a 
spectral index ranging from $\alpha\sim 1.6$ to $\alpha\sim 2$. Then the weak 
turbulence approach could explain the steepening of the spectrum. But the 
variety of physical processes in the ISM, such as shocks or dispersive effects 
for instance, do not allow to give a definitive answer but rather to ask the 
question\,: by how much is the energy spectrum of a turbulent medium affected 
by such physical processes\,? 
A formalism that incorporates dispersive effects in MHD, {\it e.g.} the Hall 
current for a strongly ionized plasma, as in the magnetosphere or in the
vicinity of proto--stellar jets or the ambipolar drift in the weakly ionized 
plasma of the interstellar medium at large, will be useful but is left for 
future work. 
So is the incorporation of compressibility.

%%%%%%%%%%%%%%%%%%%%%%%%%%%%%%%%%%%%%%%%%%%%%%%%%%%%%%%%%%%%%%%%%%%%%%%%%%%%%%%%
\section{Conclusion}
\label{swt6}
%%%%%%%%%%%%%%%%%%%%%%%%%%%%%%%%%%%%%%%%%%%%%%%%%%%%%%%%%%%%%%%%%%%%%%%%%%%%%%%%

We have obtained in this paper the kinetic equations for weak Alfv\'enic
turbulence in the presence of correlations between the velocity and the
magnetic field, and taking into account the non--mirror invariance of the MHD
equations leading to non--zero helical terms. These equations, contrary to what 
is stated in Sridhar and Goldreich (1994),
%\cite{srigo}, 
obtain at the level of three--wave interactions (see also Montgomery and 
Matthaeus 1995; Ng and Bhattacharjee 1996). 

In this anisotropic medium, a new spectral tensor must be taken into account in
the formalism when compared to the isotropic case (which can include terms
proportional to the helicity)\,; this new spectral tensor $I^s$ is linked to the 
anisotropy induced by the presence of a strong uniform magnetic field, and we 
can also study its dynamics. This purely anisotropic correlator was also 
analysed in the case of neutral fluids in the presence of rotation (Cambon and 
Jacquin 1989). 

We obtain an asymptotic two-dimensionalisation of the spectra\,: indeed, the 
evolution of the turbulent spectra at each $k_{\pa}$ is determined only by the 
spectra at the same $k_{\pa}$ and by the purely 2D state characterized by 
$k_{\pa}=0$. This property of bi--dimensionalisation was previously obtained 
theoretically from an analysis of the linearised MHD equations (Montgomery and 
Turner 1981) and using phenomenological models (Ng and Bhattacharjee 1996), and 
numerically as well (Oughton et al. 1994; Kinney and McWilliams 1998), whereas 
it is obtained in our paper 
from the rigorously derived kinetic equations. Note that the strong field 
${\bf B_0}$ has no structure (it is a ${\bf k}=0$ field), whereas the analysis 
performed in 
Kinney and McWilliams (1998) 
%\cite{kima} 
considers a strong quasi--uniform magnetic field of 
characteristic wavenumber $k_L\not= 0$, in which case the authors find that 
bi--dimensionalisation obtains as well for large enough wavenumbers.

We have considered three--dimensional turbulence in the 
asymptotic regime of large time when the spectrum tends to a quasi-2D form. 
This is the same regime for which RMHD approach is valid (Strauss 1976). However,
in addition to the shear-Alfv\'en waves described by the RMHD equations,
the kinetic equations describe also the dynamics of the so-called 
pseudo-Alfv\'en waves which are decoupled from the shear-Alfv\'en waves in 
this case, from the magnetic helicity and from the pseudo-helicity. 
Finding the Kolmogorov solution for the 3D case is technically very similar to 
the case of 2D turbulence. This leads to the spectrum 
$f(k_{\pa}) \, k_\perp^{-2}$, where $f(k_{\pa})$ is an arbitrary 
function which is to be fixed by matching in the forcing region at small 
wavenumbers. 
The $k_{\pa}$ dependence is non-universal and depends on the form of the
forcing because of the property that there is no energy transfer between 
different $k_{\pa}$'s. 

The $f(k_{\pa}) \, k_\perp^{-2}$ energy spectrum is well verified by numerics, 
which also shows that this spectrum is reached in a singular fashion with small 
scales 
developing in a finite time. We also obtain a family of Kolmogorov solutions with
different values of spectra for different wave polarities and we show that the 
sum of the spectral exponents of these spectra is equal to $-4$. The dynamics of 
both the shear-Alfv\'en waves and the pseudo-Alfv\'en waves is obtained.
Finally, the small-scale spectrum of isotropic 3D MHD turbulence in the case when 
there is no external field is also derived. 

The weak turbulence regime remains valid as long as the Alfv\'en characteristic 
time $(k_{\pa}b_0)^{-1}$ remains small compared to the transfer time (which can 
be evaluated from equation (\ref{K1})), that is to say for 
$\epsilon^2 E^s_{\bot}(k_{\bot},0) k_{\bot}^2 / B_0^2 \ll k_\pa / k_{\bot}$. 
Using the exact scaling law found in this paper, 
$E^s_{\bot}(k_{\bot},0) \propto k_{\bot}^{-2}$, we see that the condition for 
weak turbulence is less satisfied for large $k_{\bot}$, or small $k_{\pa}$. 
This means that we have a non-uniform expansion in $B_0$. 

The dynamo problem in the present formalism reduces to its simplest 
expression\,: in the presence of a strong uniform magnetic field ${\bf B_0}$, 
to a first approximation (closing the equations at the level
of second--order correlation tensors), one obtains immediate equipartition
between the kinetic and the magnetic wave energies, corresponding to
an instantaneous efficiency of the dynamo. Of course, one may ask about
the origin of ${\bf B_0}$ itself, in which case one may resort to standard
dynamo theories (see \eg Parker 1994). We see no tendency towards condensation. 

In view of the ubiquity of turbulent conducting flows embedded in strong 
quasi--uniform magnetic fields, the present derivation should be of some use
when studying the dynamics of such media, even though compressibility effects
have been ignored. It can be argued (Goldreich and Sridhar 1995)
that this incompressible approximation may be sufficient because of the 
damping of the fast magnetosonic wave by plasma kinetic effects. Note however
that Bhattacharjee, Ng and Spangler (1998) found that in the presence of spatial
inhomogeneities, there are significant departures from incompressibility at the
leading order of an asymptotic theory which assumes that the Mach number of the
turbulence is small. 
Finally, the wave energy may not remain negligible for all times, in which case
resort to phenomenological models for strong MHD turbulence is required. Is
desirable as well an exploration of such complex flows through analysis of 
laboratory and numerical experiments, and through detailed observations like 
those stemming from satellite data for the solar wind, from the THEMIS
instrument for the Sun looking at the small--scale magnetic structures of the 
photosphere, and the planned large array instrumentation (LSA--ALMA) to observe in
detail the interstellar medium.

\bigskip\bigskip\bigskip

%%%%%%%%%%%%%%%%%%%%%%%%%%%%%%%%%%%%%%%%%%%%%%%%%%%%%%%%%%%%%%%%%%%%%%%%%%%%%%%%
\section*{Acknowledgments}
This work has been performed using the computing facilities provided by 
the program ``Simulations Interactives et Visualisation en Astronomie et 
M\'ecanique (SIVAM)'' at OCA. Grants from CNRS (PNST and PCMI) and from EC
(FMRX-CT98-0175) are gratefully acknowledged.

\newpage

%%%%%%%%%%%%%%%%%%%%%%%%%%%%%%%%%%%%%%%%%%%%%%%%%%%%%%%%%%%%%%%%%%%%%%%%%%%%%%%%
\section*{Appendix}
%%%%%%%%%%%%%%%%%%%%%%%%%%%%%%%%%%%%%%%%%%%%%%%%%%%%%%%%%%%%%%%%%%%%%%%%%%%%%%%%

From the dynamical equations (\ref{eqa}) one writes successively for the second 
and third--order moments of the ${\bf z}^s$ fields\,:
\be
\begin{array}{lll}
\partial_t\left\lbrace a^s_j(\kb) a^{\sp}_{\jp}(\kbp)\right\rbrace =
&-i \epsilon k_m P_{jn}(\kb) \int  \left\lbrace
a^s_m(\kgb) a^s_n(\Lb) a^{\sp}_{\jp}(\kbp) \right\rbrace
e^{-2is \omega_{\kappa} t} \delta_{\kgb \Lb,\kb} d_{\kgb \Lb}\\
&-i \epsilon \kp_mP_{\jp m}(\kbp) \int  \left\lbrace
a^{-\sp}_m(\kgb) a^{\sp}_n({\bf L}) a^s_j(\kb) \right\rbrace
e^{-2i\sp \omega_{\kappa} t} \delta_{\kgb \Lb,\kbp} d_{\kgb \Lb}
\end{array}
\label{third}\ee
and
\be
\begin{array}{lll}
\partial_t 
\left\lbrace a^s_j(\kb) a^{\sp}_{\jp}(\kbp) a^{\ss}_{\js}(\kbs)\right\rbrace =
\end{array}
\label{fourth}
\ee
$$ -i \epsilon k_mP_{jn}(\kb) \int  \left\lbrace
a^{-s}_m(\kgb) a^s_n({\bf L}) a^{\sp}_{\jp}(\kbp) a^{\ss}_{\js}(\kbs)
\right\rbrace 
e^{-2is \omega_{\kappa} t} \delta_{\kgb \Lb,\kb} d_{\kgb \Lb}$$
$$+ \left\lbrace [\kb,s,j] \rightarrow [\kbp,\sp,\jp] 
\rightarrow [\kbs , \ss,\js] \rightarrow [\kb, s,j]
\right\rbrace \, .$$
Asymptotic closure for the leading order contributions to each of the
cumulants follows from the following procedure or algorithm. Cumulants are
in (1:1) correspondence with the moments\,: in the zero mean case, the second
and third moments are the second and third cumulants; the fourth order
moment is the sum of cumulants where each decomposition of the moment is
taken once, namely the sum of the fourth order cumulant plus products of
second order ones. One attempts to solve the hierarchy of cumulant equations
perturbatively. The asymptotic expansions generated in this way are not
uniform in time because of the presence of small divisors (mainly but not
totally due to resonances). In order to restore the uniform validity of the
asymptotic expansions we must allow the leading order contributions to each
of the cumulants vary slowly in time and choose that time dependence to
achieve that goal. Where necessary, and where the notion of well-orderedness
does not make sense in Fourier space, one must look at the corresponding
asymptotic expansions in physical space. The result is another set of
asymptotic expansions which include both the kinetic equations for the
second order moments, combinations of which give the parallel and total
energies and helicity densities, and similar equations for higher order
cumulants which can be collectively solved by a common frequency
renormalization. The kinetic equations are valid for time scales of the
order of $\epsilon^{-2}$ and possibly longer depending on how degenerate the
resonant manifold are.
Success in obtaining asymptotic closure depends on two ingredients. The
first is the degree to which the linear waves interact to randomize phases
and the second is the fact that the nonlinear regeneration of the third
order moment by the fourth order moment in equation (\ref{fourth}) depends more 
on the product of the second order moments than it does on the fourth order
cumulant. We now give details of the calculations. 

The fourth--order moment in the above equation, 
$\langle \kgb \Lb \kbp \kbs \rangle$ in short--hand notation, decomposes into 
the sum of three products of second--order moments, and a fourth--order cumulant
$\left\lbrace mn\jp\js\right\rbrace$. The latter does not contribute to secular 
behavior, and of the remaining terms one is absent as well in the kinetic 
equations because it involves the combination of wavenumbers 
$\langle \kgb \Lb \rangle \langle \kbp \kbs \rangle$\,: it introduces,
because of homogeneity, a $\delta(\kgb + \Lb)$ factor which combined with
the convolution integral leads to a zero contribution for $\kb=0$.
Hence, the time evolution of the third--order cumulants
leads to six terms that read\,:
\be
\begin{array}{lll}
&\partial_t 
\left\lbrace a^s_j(\kb) a^{\sp}_{\jp}(\kbp) a^{\ss}_{\js}(\kbs)\right\rbrace 
=-i \epsilon k_mP_{jn}(\kb) q^{-s\sp}_{m\jp}(\kbp) q^{s\ss}_{n\js}(\kbs) 
e^{2is\omp t}\\
&-i \epsilon k_mP_{jn}(\kb) q^{-s\ss}_{m\js}(\kbs) q^{s\sp}_{n\jp}(\kbp) 
e^{2is\oms t}
-i \epsilon \kp_mP_{\jp n}(\kbp) q^{-\sp \ss}_{m\js}(\kbs) q^{\sp s}_{nj}(\kb) 
e^{2i\sp \oms t}\\
&-i \epsilon \kp_mP_{\jp n}(\kbp) q^{-\sp s}_{mj}(\kb) q^{\sp \ss}_{n\js}(\kbs) 
e^{2i\sp \omega t}
-i \epsilon \ks_mP_{\js n}(\kbs) q^{-\ss s}_{mj}(\kb) q^{\ss \sp}_{n\jp}(\kbp) 
e^{2i\ss \omega t}\\
&-i \epsilon \ks_mP_{\js n}(\kbs) q^{-\ss \sp}_{m\jp}(\kbp) q^{\ss s}_{nj}(\kb) 
e^{2i\ss \omp t} \ .
\end{array}
\label{six}\ee
It can be shown that, of these six terms, only the fourth and fifth ones give
non--zero contributions to the kinetic equations.
Defining
\be
\omega_{k, \kappa L}=\omega_k-\omega_\kappa -\omega_L
\label{omn}\ee
and integrating equation (\ref{six}) over time, the exponential terms
will lead to
\be
\Delta(\omega_{k,\kappa L})=\int_0^t \exp{[it\omega_{k, \kappa L}}]dt
={{\exp{[i\omega_{k,\kappa L}]}-1}\over{i\omega_{k,\kappa L}}} \ .
\label{Delta}\ee
Substituting these expressions in (\ref{six}), only the terms 
which have an argument in the $\Delta$ functions that cancel exactly with
the arguments in the exponential appearing in (\ref{third}) will contribute.
We then obtain the fundamental kinetic equations for the energy tensor, \viz\,:
\be
\begin{array}{lll}
\partial_t q^{s\sp}_{j\jp}(\kbp) \delta(\kb+\kbp) =
\end{array}
\label{kin}
\ee
$$-{\epsilon}^2 k_mP_{jn}(\kb) \int k_{2p} P_{nq}(\Lb)
q^{-s-s}_{pm}(\kgb) q^{s \sp}_{q\jp}(\kbp) \Delta(-2s\omega_1)
\delta_{\kgb \Lb,\kb} d_{\kgb \Lb}$$
$$-{\epsilon}^2 k_mP_{jn}(\kb) \delta_{s\sp} \int \kp_{p} P_{\jp q}(\kbp)
q^{-\sp -s}_{pm}(\kgb) q^{\sp s}_{qn}(\Lb) \Delta(-2s\omega_1)
\delta_{\kgb \Lb,\kb} d_{\kgb \Lb}$$
$$-{\epsilon}^2 \kp_{m}P_{\jp n}(\kbp) \int k_{2p} P_{nq}(\Lb)
q^{-\sp -\sp}_{pm}(\kgb) q^{\sp s}_{qj}(\kb) \Delta(-2\sp \omega_1)
\delta_{\kgb \Lb,\kbp} d_{\kgb \Lb}$$
$$-{\epsilon}^2 \kp_{m}P_{\jp n}(\kbp) \delta_{s\sp} \int k_p P_{nq}(\kb)
q^{-s-\sp}_{pm}(\kgb) q^{s\sp}_{qn}(\Lb) \Delta(-2s\omega_1)
\delta_{\kgb \Lb,\kbp} d_{\kgb \Lb} \, .$$
We now perform an integration over the delta and taking the limit $t \to + \infty$
we find 
\be
\begin{array}{lll}
\partial_t[q^{s\sp}_{j\jp}(\kbp) \delta(\kb+\kbp)] =
\end{array}
\label{qssp}
\ee
$$- \epsilon^2 \int\int\int \Biggr\{Q^{-s}_k(\kgb)P_{jn}(\kb)P_{nl}(\Lb) 
[q^{s\sp}_{j\jp}(\kbp) \Bigg[\ph\delta(\kappa_{\pa})-i{\cal P}({{1}
\over{2s\kappa_{\pa}}}) \Bigg]$$
$$+Q^{-\sp}_k(\kgb)P_{\jp n}(\kb)P_{nl}(\Lb) [q^{\sp s}_{lj}(\kb)
\Bigg[\ph\delta(\kappa_{\pa})+i{\cal P}({{1}\over{2\sp\kappa_{\pa}}}) \Bigg]$$
$$- \pi \delta_{s\sp}Q^{-s}_k(\kgb)P_{\jp l}(\kb)P_{jn}(\kbp) q^{ss}_{ln}
(\Lb) \delta(\kappa_{\pa}) \Biggr\} d\kappa_1 d\kappa_2 d\kappa_{\pa} \, ,$$
where ${\cal P}$ stands for the principal value of the integral.

In the case where $s=\sp$ of interest here because the cross--correlators 
between z--fields of opposite polarities decay to zero in that
approximation, the above equations simplify to\,:
\be
\begin{array}{lll}
{{2}\over{\pi}}\partial_t [q^{ss}_{j\jp}(\kbp) \pm q^{ss}_{\jp j}(\kbp) ] =
\end{array}
\label{qss}
\ee
$$2 \epsilon^2 \int P_{jn}(\kb)P_{\jp q}(\kb) [q^{ss}_{qn}(\Lb)\pm q^{ss}_{nq}(\Lb)]
Q^{-s}_k(\kgb) \delta(\kappa_{\pa}) d\kappa_1 d\kappa_2 d\kappa_{\pa}$$
$$-\epsilon^2 \int P_{jn}(\kb)P_{nq}(\Lb) [q^{ss}_{q\jp}(\kbp)\pm q^{ss}_{\jp q}
(\kbp)] Q^{-s}_k(\kgb) \delta(\kappa_{\pa}) d\kappa_1 d\kappa_2 d\kappa_{\pa}$$
$$-\epsilon^2 \int P_{\jp n}(\kb)P_{nq}(\Lb) [q^{ss}_{jq}(\kbp)\pm q^{ss}_{qj}
(\kbp)] Q^{-s}_k(\kgb) \delta(\kappa_{\pa}) d\kappa_1 d\kappa_2 d\kappa_{\pa}$$
$$+{i \epsilon^2 \over s\pi} {\cal P} 
\int P_{jn}(\kb)P_{nq}(\Lb) [q^{ss}_{q\jp}(\kbp)\mp q^{ss}_{\jp q}(\kbp)]
Q^{-s}_k(\kgb) {{d\kappa_1 d\kappa_2 d\kappa_{\pa}}\over{\kappa_{\pa}}}$$
$$-{i \epsilon^2 \over s\pi} {\cal P} 
\int P_{\jp n}(\kb)P_{nq}(\Lb) [q^{ss}_{jq}(\kbp)\mp q^{ss}_{qj}(\kbp)]
Q^{-s}_k(\kgb) {{d\kappa_1 d\kappa_2 d\kappa_{\pa}}\over{\kappa_{\pa}}} \, .$$

To derive the kinetic equations we need now to develop 
\be
\begin{array}{lll}
\partial_t e^s(\kb) = \partial_t ( q^{ss}_{11}(\kb) + q^{ss}_{22}(\kb) + 
q^{ss}_{{\pa}\,{\pa}}(\kb) ) \ , 
\end{array}
\ee
$$\partial_t \Phi^s(\kb) = \kbpe^{-4} \partial_t q^{ss}_{{\pa}\,{\pa}}(\kb) \ ,$$
$$\partial_t R^s(\kb) = {1 \over -i k_{\pa} \kbpe^2} \partial_t 
( q^{ss}_{12}(\kb) - q^{ss}_{21}(\kb) ) \ ,$$
$$\partial_t  I^s(\kb) = \kbpe^{-4} \partial_t ( k_2( q^{ss}_{1{\pa}}(\kb) + 
q^{ss}_{{\pa}1}(\kb) ) - k_1( q^{ss}_{2{\pa}}(\kb) + q^{ss}_{{\pa}2}(\kb) )) \ ,$$
in terms of the above expressions (\ref{qssp}) and (\ref{qss}). This leads to\,:

\be
\begin{array}{lll}
\partial_t e^s(\kb) = 
\end{array}
\label{ab2}
\ee
$$
{\pi \epsilon^2 \over 2} \int [ 2( q^{ss}_{11}(\Lb) + 
q^{ss}_{22}(\Lb) + q^{ss}_{{\pa}\,{\pa}}(\Lb) - q^{ss}_{11}(\kb) - 
q^{ss}_{22}(\kb) + q^{ss}_{{\pa}\,{\pa}}(\kb) )$$
$$+ {L_nL_l \over L^2} (q^{ss}_{ln}(\kb) + q^{ss}_{nl}(\kb)) - {k_nk_l \over k^2} 
(q^{ss}_{ln}(\Lb) + q^{ss}_{nl}(\Lb)) ] \, Q^{-s}_k(\kgb) \delta(\kappa_{\pa}) 
d\kappa_1 d\kappa_2 d\kappa_{\pa} \ ,$$

\newpage

\be
\begin{array}{lll}
\partial_t \Phi^s(\kb) =
\end{array}
\label{ab3}
\ee
$${\pi \epsilon^2 \over 2 \kbpe^4} \int [2 
q^{ss}_{{\pa}\,{\pa}}(\Lb) - 2{k_{\pa} k_l \over \kb^2} 
( q^{ss}_{l{\pa}}(\Lb) + q^{ss}_{{\pa}l}(\Lb) ) + {k_{\pa}^2 \over \kb^4} k_nk_l
( q^{ss}_{ln}(\Lb) + q^{ss}_{nl}(\Lb) )]$$
$$+ [-2 q^{ss}_{{\pa}\,{\pa}}(\kb) + ( {L_{\pa}L_l \over L^2} + {k_{\pa}k_l \over 
k^2} - {k_{\pa} \kb \cdot \Lb \over \kb^2 \Lb^2} L_l ) 
( q^{ss}_{l{\pa}}(\kb) + q^{ss}_{{\pa}l}(\kb) ) ] \, Q^{-s}_k(\kgb) 
\delta(\kappa_{\pa}) d\kappa_1 d\kappa_2 d\kappa_{\pa} $$
$$+ {i \epsilon^2 s \over 2} {\cal P} \int [ -( {L_{\pa}L_l \over L^2} + 
{k_{\pa}k_l \over k^2} - {k_{\pa} \kb \cdot \Lb \over \kb^2 \Lb^2} L_l )
( q^{ss}_{l{\pa}}(\kb) - q^{ss}_{{\pa}l}(\kb) ) ] \, Q^{-s}_k(\kgb) 
{{d\kappa_1 d\kappa_2 d\kappa_{\pa}}\over{\kappa_{\pa}}} \ ,
$$

\vspace{.4cm}

\be
\begin{array}{lll}
\partial_t R^s(\kb) =
\end{array}
\label{ab4}
\ee
$${i \pi \epsilon^2 \over 2 k_{\pa} \kpe^2} [ 2 \int 
( q^{ss}_{21}(\Lb) - q^{ss}_{12}(\Lb) ) - {k_2 k_l \over \kb^2} 
( q^{ss}_{l1}(\Lb) - q^{ss}_{1l}(\Lb) ) - {k_1 k_n \over \kb^2} 
( q^{ss}_{2n}(\Lb) - q^{ss}_{n2}(\Lb) )$$ 
$$+ {k_1 k_2 k_n k_l\over \kb^4} ( q^{ss}_{ln}(\Lb) - q^{ss}_{nl}(\Lb) ) \, 
Q^{-s}_k(\kgb) \delta(\kappa_{\pa}) d\kappa_1 d\kappa_2 d\kappa_{\pa}$$
$$- \int ( q^{ss}_{12}(\kb) - q^{ss}_{21}(\kb) ) - {L_1 L_l \over \Lb^2} 
( q^{ss}_{l2}(\kb) - q^{ss}_{2l}(\kb) ) - {k_1 k_l \over \kb^2} 
( q^{ss}_{l2}(\kb) - q^{ss}_{2l}(\kb) )$$ 
$$+ {k_1 \kb \cdot \Lb L_l \over \kb^2 \Lb^2} ( q^{ss}_{l2}(\kb) - 
q^{ss}_{2l}(\kb) ) \, Q^{-s}_k(\kgb) \delta(\kappa_{\pa}) 
d\kappa_1 d\kappa_2 d\kappa_{\pa} $$
$$- \int ( q^{ss}_{12}(\kb) - q^{ss}_{21}(\kb) ) - {L_2 L_l \over \Lb^2} 
( q^{ss}_{1l}(\kb) - q^{ss}_{l1}(\kb) ) - {k_2 k_l \over \kb^2} 
( q^{ss}_{1l}(\kb) - q^{ss}_{l1}(\kb) )$$
$$ + {k_2 \kb \cdot \Lb L_l \over \kb^2 \Lb^2} ( q^{ss}_{1l}(\kb) - 
q^{ss}_{l1}(\kb) ) \, Q^{-s}_k(\kgb) \delta(\kappa_{\pa}) 
d\kappa_1 d\kappa_2 d\kappa_{\pa} $$
$$+{i s {\cal P} \over \pi} \int [ - {L_1 L_l \over \Lb^2} 
( q^{ss}_{l2}(\kb) + q^{ss}_{2l}(\kb) ) - {k_1 k_l \over \kb^2} 
( q^{ss}_{l2}(\kb) + q^{ss}_{2l}(\kb) )$$ 
$$+ {k_1 \kb \cdot \Lb L_l \over \kb^2 
\Lb^2} ( q^{ss}_{l2}(\kb) + q^{ss}_{2l}(\kb) ) + {L_2 L_l \over \Lb^2} 
( q^{ss}_{1l}(\kb) + q^{ss}_{l1}(\kb) ) + {k_2 k_l \over \kb^2} 
( q^{ss}_{1l}(\kb) + q^{ss}_{l1}(\kb) )$$
$$ - {k_2 \kb \cdot \Lb L_l \over 
\kb^2 \Lb^2} ( q^{ss}_{1l}(\kb) + q^{ss}_{l1}(\kb) )] \, Q^{-s}_k(\kgb) 
{{d\kappa_1 d\kappa_2 d\kappa_{\pa}}\over{\kappa_{\pa}}} \, ,$$

\newpage

\be
\begin{array}{lll}
\partial_t I^s(\kb) =
\end{array}
\label{ab5}
\ee
$${\pi \epsilon^2 k_2 \over 2 \kpe^4} [ 2 \int 
( q^{ss}_{1 \pa}(\Lb) + q^{ss}_{\pa 1}(\Lb) ) - {k_1 k_l \over \kb^2} 
( q^{ss}_{l \pa}(\Lb) + q^{ss}_{\pa l}(\Lb) ) - {k_{\pa} k_n \over \kb^2} 
( q^{ss}_{1n}(\Lb) + q^{ss}_{n1}(\Lb) )$$ 
$$+ {k_{\pa} k_1 k_n k_l \over \kb^4} ( q^{ss}_{ln}(\Lb) + q^{ss}_{nl}(\Lb) ) 
\, Q^{-s}_k(\kgb) \delta(\kappa_{\pa}) d\kappa_1 d\kappa_2 d\kappa_{\pa}$$
$$- \int ( q^{ss}_{\pa 1}(\kb) + q^{ss}_{1 \pa}(\kb) ) + ( -{L_{\pa} L_l \over 
\Lb^2} - {k_{\pa} k_l \over \kb^2} + { k_{\pa} \kb \cdot \Lb L_l \over \kb^2 
\Lb^2} ) ( q^{ss}_{l1}(\kb) + q^{ss}_{1l}(\kb) ) \, Q^{-s}_k(\kgb) 
\delta(\kappa_{\pa}) d\kappa_1 d\kappa_2 d\kappa_{\pa} $$
$$- \int ( q^{ss}_{\pa 1}(\kb) + q^{ss}_{1 \pa}(\kb) ) + (-{L_1 L_l \over \Lb^2}
- {k_1 k_l \over \kb^2} + { k_1 \kb \cdot \Lb L_l \over \kb^2 \Lb^2} )
( q^{ss}_{\pa l}(\kb) + q^{ss}_{l \pa}(\kb) ) \, 
Q^{-s}_k(\kgb) \delta(\kappa_{\pa}) d\kappa_1 d\kappa_2 d\kappa_{\pa}$$
$$+{i s {\cal P} \over \pi} \int 
( -{L_{\pa} L_l \over \Lb^2} - {k_{\pa} k_l \over \kb^2} + { k_{\pa} \kb 
\cdot \Lb L_l \over \kb^2 \Lb^2} ) ( q^{ss}_{l1}(\kb) - q^{ss}_{1l}(\kb) )$$
$$+(-{L_1 L_l \over \Lb^2} - {k_1 k_l \over \kb^2} + { k_1 \kb \cdot \Lb L_l 
\over \kb^2 \Lb^2} ) ( q^{ss}_{\pa l}(\kb) - q^{ss}_{l \pa}(\kb) ) \, 
Q^{-s}_k(\kgb) {{d\kappa_1 d\kappa_2 d\kappa_{\pa}}\over{\kappa_{\pa}}}]$$
$$-{\pi \epsilon^2 k_1 \over 2 \kpe^4} [ 2 \int 
( q^{ss}_{\pa 2}(\Lb) + q^{ss}_{2 \pa}(\Lb) ) - {k_{\pa} k_l \over \kb^2} 
( q^{ss}_{l 2}(\Lb) + q^{ss}_{2 l}(\Lb) ) - {k_2 k_n \over \kb^2} 
( q^{ss}_{\pa n}(\Lb) + q^{ss}_{n \pa}(\Lb) )$$
$$+ {k_2 k_{\pa} k_n k_l \over \kb^4} ( q^{ss}_{ln}(\Lb) + q^{ss}_{nl}(\Lb) ) 
\, Q^{-s}_k(\kgb) \delta(\kappa_{\pa}) d\kappa_1 d\kappa_2 d\kappa_{\pa}$$
$$- \int ( -{L_2 L_l \over \Lb^2} - {k_2 k_l \over \kb^2} + 
{ k_2 \kb \cdot \Lb L_l \over \kb^2 \Lb^2} ) ( q^{ss}_{l \pa}(\kb) + 
q^{ss}_{\pa l}(\kb) )$$
$$ + (-{L_{\pa} L_l \over \Lb^2} - {k_{\pa} k_l \over \kb^2} + 
{ k_{\pa} \kb \cdot \Lb L_l \over \kb^2 \Lb^2} ) ( q^{ss}_{2l}(\kb) + 
q^{ss}_{l2}(\kb) ) \, 
Q^{-s}_k(\kgb) \delta(\kappa_{\pa}) d\kappa_1 d\kappa_2 d\kappa_{\pa}$$
$$+{i s {\cal P} \over \pi} \int 
( -{L_2 L_l \over \Lb^2} - {k_2 k_l \over \kb^2} + { k_2 \kb 
\cdot \Lb L_l \over \kb^2 \Lb^2} ) ( q^{ss}_{l \pa}(\kb) - q^{ss}_{\pa l}(\kb) )$$
$$+(-{L_{\pa} L_l \over \Lb^2} - {k_{\pa} k_l \over \kb^2} + { k_{\pa} \kb 
\cdot \Lb L_l \over \kb^2 \Lb^2} ) ( q^{ss}_{2l}(\kb) - q^{ss}_{l2}(\kb) ) \, 
Q^{-s}_k(\kgb) {{d\kappa_1 d\kappa_2 d\kappa_{\pa}}\over{\kappa_{\pa}}}] \, .$$

The final step which leads to the kinetic equations (\ref{K1})--(\ref{K4}), 
consists in substituting the expressions (\ref{corrq}) inside 
(\ref{ab2})--(\ref{ab5}). For this computation it is useful to note that\,: 
\be
\begin{array}{lll}
X^2 + Y^2 = \kappa_{\perp}^2 k_{\perp}^2 \ ,
\end{array}
\ee
$$X^2 + Z^2 = k_{\perp}^2 L_{\perp}^2 \ ,$$
$$Z^2 - X^2 = (L_1^2-L_2^2)(k_1^2-k_2^2) + 4k_1 k_2 L_1 L_2 \ ,$$
$$XZ = L_1L_2(k_2^2-k_1^2)+k_1k_2(L_1^2-L_2^2) \ ,$$
$$XY = k_{\perp}^2(k_2L_1-k_1L_2)+L_1L_2(k_1^2-k_2^2)+k_1k_2(L_2^2-L_1^2) \ .$$

\vfill\eject

%%%%%%%%%%%%%%%%%%%%%%%%%%%%%%%%%%%%%%%%%%%%%%%%%%%%%%%%%%%%%%%%%%%%%%%%%%%%%%%%

\vfill\eject

\vfill\eject

\vfill\eject

\vfill\eject

\vfill\eject

\vfill\eject

\vfill\eject

\vfill\eject

\vfill\eject

\vfill\eject

\vfill\eject

\vfill\eject

\end{document}